\documentclass[acmsmall,screen]{acmart}
\usepackage{algorithm}
\usepackage{cleveref}
\usepackage{makecell}
\usepackage{xspace}
\usepackage{soul}
\usepackage{sidecap}
\usepackage{algpseudocode}

\newcommand{\bound}{\ensuremath{\operatorname{bound}}\xspace}
\newcommand{\ampl}{\ensuremath{\operatorname{ampl}}\xspace}
\newcommand{\intro}{\ensuremath{\operatorname{intro}}\xspace}
\newcommand{\lo}{\ensuremath{\operatorname{lo}}\xspace}
\newcommand{\hi}{\ensuremath{\operatorname{hi}}\xspace}
\newcommand{\maxlog}{\ensuremath{\operatorname{maxlog}}\xspace}
\newcommand{\minlog}{\ensuremath{\operatorname{minlog}}\xspace}
\newcommand{\logspan}{\ensuremath{\operatorname{logspan}}\xspace}
\newcommand{\slack}{\ensuremath{\operatorname{slack}}\xspace}
\newcommand{\pow}{\ensuremath{\operatorname{pow}}\xspace}
\newcommand{\atan}{\ensuremath{\operatorname{atan}}\xspace}

\newcommand{\toolname}{Reval\xspace}
\newcommand{\RivalHerbieAverageSpeedup}{1.37}

\newcommand{\NumBenchmarks}{493\xspace}
\newcommand{\NumBenchmarksDiscardedbyFunctionality}{14\xspace}

\newcommand{\NumBenchmarksIncludeDiscarded}{509\xspace}

\newcommand{\NumBenchOpsMin}{1\xspace}
\newcommand{\NumBenchOpsMax}{59\xspace}
\newcommand{\NumBenchOpsAvg}{11.6\xspace}

\newcommand{\NumBenchVarsMin}{1\xspace}
\newcommand{\NumBenchVarsMax}{9\xspace}
\newcommand{\NumBenchVarsAvg}{2.5\xspace}

\newcommand{\NumPointsPerBenchmark}{8256\xspace}

\newcommand{\NumTunedBenchmarks}{482\xspace}

\newcommand{\NumTunedPoints}{567\thinspace459\xspace}

\newcommand{\RivalAvgSpeedupOverSollya}{3.97\xspace}

\newcommand{\RivalMaxSpeedupOverSollya}{9.26\xspace}

\newcommand{\SollyaFaithfulCnt}{5357\xspace}

\newcommand{\TuningTimePercentage}{2.6}

\newcommand{\RivalSecondIterConvergence}{59.05}
\newcommand{\BaselineSecondIterConvergence}{32.38}

\newcommand{\RivalInstrCountLessThanBaseline}{31.7}
\newcommand{\RivalInstrCountLessThanZiv}{62.16}
\newcommand{\DensityPercentageOfLowerPrecisionReval}{43.64}
\newcommand{\DensityPercentageOfLowerPrecisionBaseline}{31.93}
\newcommand{\DensityPercentageOfLowerPrecisionZiv}{1.06}
\newcommand{\DensityPercentageOfLowerPrecisionOptimal}{50.41}
\newcommand{\DensityAdvantageRevalOverBaseline}{1.37}
\newcommand{\DensityAdvantageRevalOverZiv}{41.17}
\newcommand{\RevalCloserToOptimalThanZiv}{14.41}
\newcommand{\RevalCloserToOptimalThanBaseline}{8.01}

\begin{document}

\title{Fast Mixed-Precision Real Evaluation}

\author{Artem Yadrov}
\affiliation{
  \institution{University of Utah}
  \city{Salt Lake City}
  \state{UT}
  \country{USA}}
\orcid{0000000236991518}
\email{artemya@cs.utah.edu}

\author{Aditya Kumar}
\affiliation{
  \institution{University of Washington}
  \city{Seattle}
  \state{WA}
  \country{USA}}
\orcid{0009000432669160}
\email{akum@cs.washington.edu}

\author{Pavel Panchekha}
\affiliation{
  \institution{University of Utah}
  \city{Salt Lake City}
  \state{UT}
  \country{USA}}
\orcid{0000000326213592}
\email{pavpan@cs.utah.edu}

\begin{abstract}
  Evaluating real numbers to high precision
    is important in computational science, numerical computing,
    and in hardware and software design.
  Efficient algorithms exist for computing
    arithmetic operations and common elementary functions,
    but composing these primitives into larger expressions
    remains a challenge.
  The typical expression-evaluation algorithm
    computes all primitives in the expression
    to a uniform precision,
    doubling that precision
    until a sufficiently-accurate result is achieved.
  This is wasteful:
    usually only a few operations
    really need to be performed at high precision,
    and the bulk of the expression
    could be computed at a much lower, faster precision.
  However, no rigorous algorithm currently exists
    to compute these non-uniform precision assignments,
    nor to leverage non-uniform precision
    to reduce the number of primitive computations needed
    to evaluate an expression.

  We propose a new algorithm
    for deriving such rigorous precision assignments
    and reducing the number of primitive computations needed.
  This algorithm can leverage results
    computed at lower precisions
    to analytically determine a precision for each primitive
    that will result in a sufficiently-accurate evaluation.
  It can also use information from constant subexpressions,
    movability flags, and repeat flags
    to skip the computation of certain primitives,
    further reducing the time to evaluate real expressions.
  Our implementation, \toolname,
    achieves an average speed-up of
    $\RivalAvgSpeedupOverSollya\times$
    compared to the widely-used Sollya tool,
    with the speed-up increasing to
    $\RivalMaxSpeedupOverSollya\times$
    on the most difficult inputs.
  A detailed evaluation shows
    that this speed-up results from both
    assigning lower precisions to the majority of primitives
    and from computing fewer primitives overall.
\end{abstract}

\maketitle


\section{Introduction}
\label{sec:introduction}
% Problem in the real world

Computational number theory~\cite{flint,mpfun}, 
  physics~\cite{bailey-physics}, 
  geometry~\cite{cgal,stoneworks}, 
  and commutative algebra~\cite{macaulay2}
  all require evaluating real numbers to high precision.
In numerical computing,
  high-precision real evaluation is used
  to fit polynomial approximations%
  ~\cite{chebfun,remez,rlibm}
  and compute data for table-based implementations%
  ~\cite{muller-handbook, muller-elementary}.
In hardware design,
  real evaluation forms a trusted, high-precision ``golden model''
  for testing and evaluation~\cite{metalibm, crlibm}.
For this reason,
  experts have developed specialized algorithms~\cite{mpfr}
  for high-precision arithmetic operators
  and common elementary functions.

However, AI has dramatically broadened the range
  of important mathematical operations to include
  activation functions like SwiGLU~\cite{swiglu}
  and vector transcendentals like softmax~\cite{softmax}.
New operations are proposed at a rapid clip,
  and no algorithms for evaluating them to high precision
  currently exist.
Composing existing high-precision primitives
  is an attractive solution, but an algorithm is needed
  to determine the precision to compute each primitive at.
Since some of the new operations
  are numerically unstable~\cite{deepstability},
  determining this precision is not straightforward.
Nor is this problem confined to machine learning;
  numerical compilers~\cite{herbie,movability}
  and high-level implementation languages%
  ~\cite{megalibm,stoneworks}
  also need similar capabilities.

The problem is easy to state.
Given an expression composed of primitives,
  concrete values for all variables in the expression,
  and a target precision,
  compute the real result of the expression
  to that target precision.
The standard approach
  is to use high-precision \emph{interval arithmetic}%
  ~\cite{sollya,herbie,arb},
  which can bound the effect of rounding error.
Specifically, Ziv's ``onion-peeling'' method~\cite{ziv}
  evaluates each primitive at some precision
  and checks if the resulting interval is narrower
  than the chosen precision bound.
If it is, then either endpoint of the interval
  is a valid result.

If it's not, one must re-evaluate at a higher precision.
Typically, one doubles it,
  but this is widely understood to be sub-par.
A comment in the Marshall interpreter~\cite{stoneworks},
  for example, suggests that
  ``we should do something more intelligent about that
  (not all subexpressions of $e$ need the same precision)'',
  while the Sollya language~\cite{sollya} implementation
  describes the choice of precision as a ``matter of religion''
  with doubling being ``fine in practice.''
The issue is that using the same precision for all primitives
  means using a precision that is too high
  for all but the most precision-sensitive operations,
  wasting valuable time.
Nonetheless, no better, practical method exists.

We provide one in \toolname, a new library for
  \textbf{r}eal \textbf{e}valuation using inter\textbf{val}s.%
\footnote{The name is also a pun on Rival,
  the underlying interval library \toolname uses.}
\toolname uses a \emph{precision assignment algorithm}
  to determine what precision to use
  for every primitive in the expression.
Critically, each primitive gets its own, often low, precision,
  leading to a dramatic acceleration of Ziv's algorithm.
\toolname's approach also has secondary benefits.
The precision assignment can skip recomputing primitives
  when their precision assignment doesn't change,
  and can leverage additional metadata like
  movability flags~\cite{movability}
  to further improve the precision assignment.
\toolname's precision assignments come
  with rigorous guarantees
  drawn from the error analysis literature%
  ~\cite{fptaylor,satire,atomu}.
Moreover, the precision assignments can be computed quickly.

We evaluate \toolname on
  \NumTunedPoints computations
  drawn from \NumBenchmarks benchmarks
  in the Herbie~2.0 benchmark suite,
  and compare it to
  the state-of-the-art Sollya tool~\cite{sollya}
  as well as to ablations of \toolname.
\toolname is $\RivalAvgSpeedupOverSollya\times$ faster
  than Sollya with speedups of
  up to $\RivalMaxSpeedupOverSollya\times$
  on the most difficult evaluations.
We show that this performance advantage
  is due to \toolname's
  highly non-uniform precision assignments
  and fewer operator evaluations.
A comparison to an offline ``oracular'' precision assignment
  shows that, on these computations,
  \toolname's precision assignments are
  $\RevalCloserToOptimalThanZiv\times$ closer to optimal
  than the traditional algorithm's.
To demonstrate real-world applicability,
  we integrate \toolname into the Herbie numerical compiler,
  replacing an earlier implementation based on Ziv's strategy.
The integration achieves
  a speedup of $\RivalHerbieAverageSpeedup\times$,
  showing that \toolname's advantages are relevant
  in practical, widely-used tools.

\medskip
\noindent
In summary, this paper contributes:
\begin{itemize}
\item
A mathematical foundation for precision assignment
in expression evaluation (\Cref{sec:bounds}).
\item
A precision assignment algorithm
based on that foundation (\Cref{sec:machine}).
\item
A practical implementation thereof,
including a number of optimizations (\Cref{sec:opt}).
\end{itemize}

\section{Overview}
\label{sec:overview}
Consider the task of evaluating the expression
  $(1.0 - \cos(x)) / \sin(x)$
  for $x = 10^{-8}$.
For this input, the subtraction at $1 - \cos(x)$
  suffers from a numerical problem known as
  ``catastrophic cancellation'',
  so ordinary floating-point arithmetic is not accurate.
A high-precision evaluation is necessary;
  Ziv's strategy is the standard approach.
This strategy suggests we evaluate the expression
  using arbitrary-precision interval arithmetic.
For example, we might choose to use 64-bit intervals,
  in which case the expression evaluates to the interval
  $[4.998\cdot10^{-9}, 5.004\cdot10^{-9}]$.
If we only need the first few digits of the result,
  this is good enough,
  but if we need more than 3 digits of accuracy,
  this interval is too wide.
Ziv's strategy then recommends trying again
  with higher precision;
  typically one doubles the precision.
With 128~bits, then,
  the expression evaluates to a very narrow interval
\[
[5.0000000000000001\ldots\cdot10^{-9},
 5.0000000000000001\ldots\cdot10^{-9}].
\]
Here both endpoints round to the same
  double-precision number, $5 \cdot 10^{-9}$,
  which we then know is the correct double-precision result.

\begin{figure}
    \includegraphics[width=.6\linewidth,trim={
      1in 5.6in 12.75in .25in
    },clip]{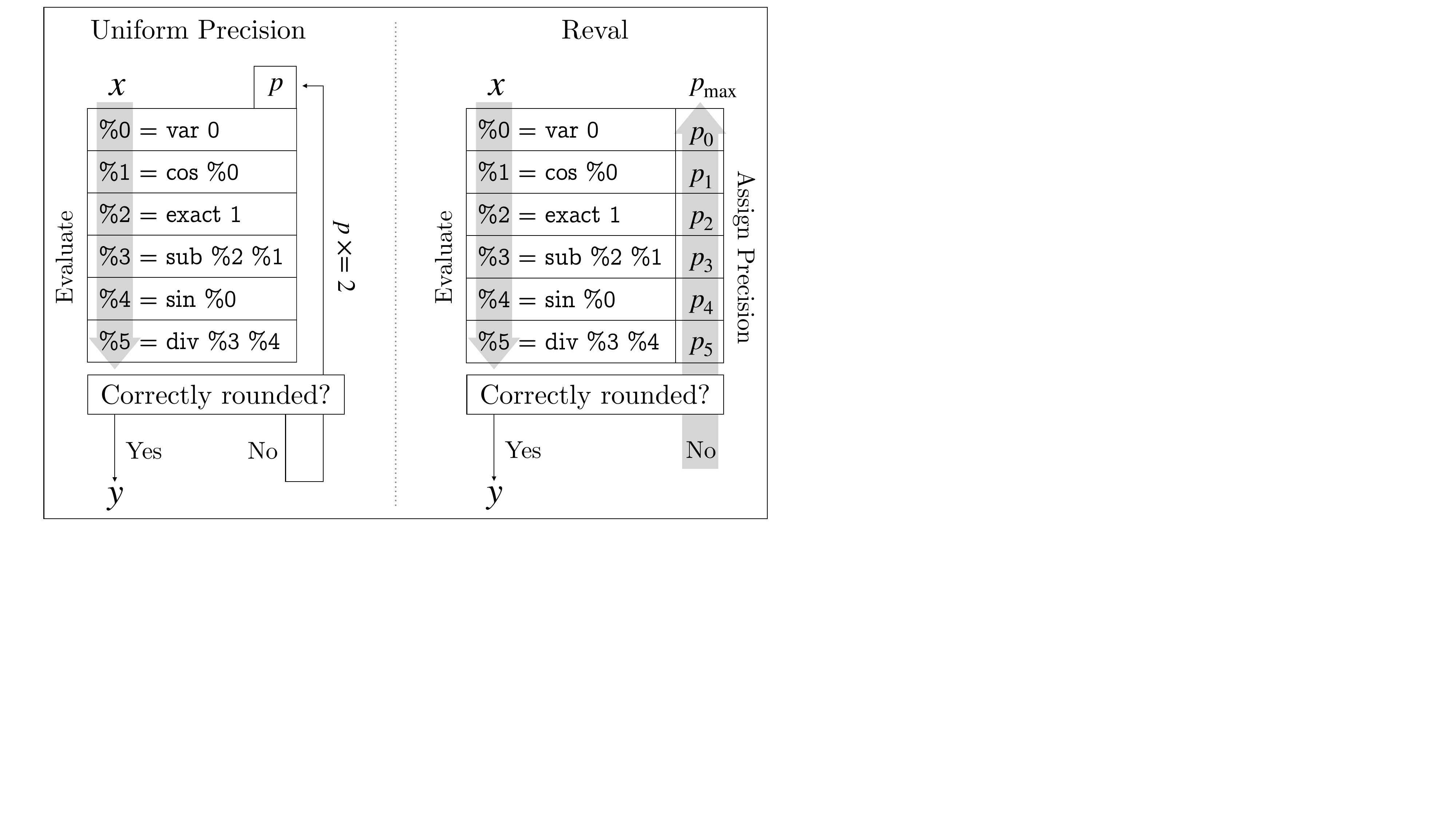}
    \caption{
    In the standard uniform-precision approach (left),
      all operations share the same precision,
      which is doubled until a correct rounding is found.
    In \toolname (right),
      precision tuning is performed
      if a correct rounding is not found,
      deriving a different precision for each operation.
    }
    \label{fig:diagram}
\end{figure}

Ziv's strategy is widely-used and efficient,
  but it leaves the actual choice of precision unspecified,
  and in fact there are many possible
  precision choices that work.
Trying 64 and then 128 bits is the standard approach---%
  doubling every time evaluation fails---%
  but here 128~bits aren't necessary;
  one can use as few as 107 bits of precision.
And the precision also need not be uniform!
For this expression, only the $\cos(x)$ term
  actually needs to be computed with 107~bits;
  every other operation can actually use
  as few as 56~bits and still produce an interval
  that identifies the correct double-precision result.

And picking the right precision is important,
  because larger precisions take more time.
In this example, the precisions are quite small
  but already the non-uniform precision is 6\% faster
  than the uniform one.
The gap is larger when larger precisions are needed;
  for example, if the input were instead $x = 10^{-80}$,
  then uniform 1024-bit precision would take 55 microseconds,
  uniform 585-bit precision would take 20 microseconds,
  and computing only the $\cos(x)$ term with 585 bits,
  with the rest of the expression using 60 bits,
  takes only 5 microseconds, a dramatic speedup.
Choosing the right precisions is critical.

That is exactly what \toolname does.
In the case of $x = 10^{-8}$, \toolname assigns
  117~bits to $\cos(x)$ and 58 or 60 bits to all other operations,
  which is only 1\% slower than
  the optimal precision assignment.
For $x = 10^{-80}$, \toolname assigns
  622~bits to $\cos(x)$ and 58 or 60 bits to all other operations,
  which is again only 3\% slower than
  the optimal precision assignment.
Moreover,
  \toolname computes these
  non-uniform precision assignments quickly---%
  about seven microseconds.
It does this through a combination
  of analytically computing
  the correct precisions for most operations
  plus an iterative guessing approach
  for operations that cannot be analyzed analytically,
  shown in \Cref{fig:diagram}.

Consider first the easier task
  of evaluating our expression at $x = 10^{-8}$.
Here, \toolname first tries to evaluate the expression
  at low precisions (between 58~and 62~bits).
Even though the output interval isn't accurate enough,
  this evaluation isn't totally useless:
  every interval computed by each operation
  contains useful information about what precision
  that operation needs to be evaluated at.
For example the expression $(1.0 - \cos(x)) / \sin(x)$
  initially evaluates to 
  $[4.987\times10^{-9}, 5.009\times10^{-9}]$.
The endpoints of this interval have
  a relative error of about $2^{-8}$,
  meaning that we have already computed
  something like 8 correct bits.
Since we want a final output with 53 bits of precision,
  this suggests adding about 45 bits of precision.
The general insight is that,
  even when the assigned precision is too low,
  the intervals computed for intermediate values
  can inform the precision assignment
  for later iterations.

More formally, \toolname computes its precisions
  by computing, from each intermediate interval $\hat{z}$,
  $\intro(\hat{z})$ and $\ampl_k(\hat{z})$ values
  that are then added to get the precision assignment
  for each operation.
For our example,
  the precision assigned to the $\cos(x)$ term is
  computed with
\[
  t +
  \lceil\log_2\ampl_1(\text{div})\rceil + 2 +
  \lceil\log_2\ampl_2(\text{sub})\rceil + 2 +
  \lceil\log_2\intro(\text{cos})\rceil + 2 + 3,
\]
  where $t$ is the target precision,
  53 bits (meaning double precision).
\toolname evaluates $\lceil\log_2\ampl_1(\text{div})\rceil = 0$,
  $\lceil\log_2\ampl_2(\text{sub})\rceil = 55$,
  $\lceil\log_2\intro(\text{cos})\rceil = 0$,
  leading to a final precision assignment of 117 bits
  for the $\cos(x)$ term.
These $\intro$ and $\ampl_k$ terms
  are defined in \Cref{sec:bounds}.

Evaluating our example expression at, let's say $x = 10^{-18}$,
  is a little harder.
In this case, the initial evaluation of $1 - \cos(x)$
  yields the interval $[0, 2.16\cdot10^{-19}]$.
The endpoints of this interval have
  \emph{infinite} relative error,
  so it doesn't provide \toolname
  with much useful information about what precision
  to evaluate $\cos(x)$ at.
\toolname falls back to Ziv-style guesses,
  and chooses to evaluate $\cos(x)$
  at 178 bits of precision
  (\Cref{sec:machine} explains how the guess is chosen).
However, even though $\cos(x)$'s precision is a guess,
  all other operations have small relative errors,
  so \toolname keeps other precision assignments low.
The resulting precision assignment---%
  178 bits for $\cos(x)$
  and 58 or 60 bits for all other operations---%
  does produce a narrow-enough interval,
  meaning that \toolname's guess was successful.
Not only is the resulting assignment close to optimal,
  but \toolname also finds it
  in only two evaluations of the expression,
  whereas the standard approach would need three evaluations
  to find a working precision.

Of course, in this example the guess worked---%
  perhaps the reader does not want to credit
  \toolname with correct \emph{guesses}.
But the same precision-assignment process
  continues to help
  even when a guess fails.
For example, on the input $x = 10^{-26}$,
  \toolname's second evaluation,
  which uses 178~bits for $\cos(x)$,
  still fails to produce a narrow-enough interval.
However, the interval for $1 - \cos(x)$
  has now narrowed to
  $[4.95\cdot10^{-53}, 5.22\cdot10^{-53}]$,
  so it no longer has unbounded relative error.
This allows \toolname to assign a new precision
  of 236~bits for $\cos(x)$.
This precision assignment finally succeeds.
Even with one failed guess,
  \toolname still needs only three total evaluations,
  while the standard approach needs four.
Moreover, all three evaluations assign the same precision
  to the $\sin(x)$ term,
  so \toolname computes it only once.
By contrast, the standard approach increases the precision
  of every operation at each re-evaluation,
  and therefore must repeatedly reevaluate this expensive term.

To summarize,
  \toolname initially performs an evaluation at low precision.
It then uses the intervals computed during that initial evaluation
  to compute $\intro$ and $\ampl_k$ factors,
  which are then used to compute a precision assignment
  for each operation.
Typically, this precision assignment is computed analytically,
  in which case re-evaluating with that precision
  computes an accurate result.
In some cases, \toolname is forced to guess;
  in those cases, re-evaluating with the computed precisions
  might still not be accurate enough,
  because the guess might be bad.
Still, \toolname can then repeat the precision assignment
  using the new,
  and hopefully narrower and more useful, intervals.

The rest of this paper is organized as follows:
  \Cref{sec:background} covers
  floating-point and interval arithmetic in brief.
\Cref{sec:bounds} then derives
  the $\intro$ and $\ampl_k$ factors and shows
  how they relate to precision tuning.
\Cref{sec:machine} then describes
  the ``slack mechanism'' by which
  \toolname guesses precisions when forced,
  and \Cref{sec:opt} describes additional optimizations
  enabled by \toolname's non-uniform precision assignments.

\section{Background}
\label{sec:background}
A normal floating-point number
  is a number $\pm (1 + m) 2^e$
  where the mantissa $m$ is a value in $[0, 1)$
  represented with some fixed precision $p \ge 2$
  and where the exponent $e$ is a signed integer.
Rival uses the well-known MPFR library~\cite{mpfr},
  where the precision is user-selected up to thousands of bits,
  the exponent is a signed 32-bit integer,
  and subnormals are not present.%
\footnote{MPFR can emulate subnormals
  by truncating and shifting,
  but its internal data representation doesn't use them.}

A real number $x$ can be \textit{rounded}
  to a precision-$p$ floating-point value.
Typically, $x$ lies strictly between
  two floating-point values,
  in which case we say that either one
  is a \textit{faithful rounding} $\hat{x}$ of $x$.
A specific \textit{rounding mode},
  such as the default round-to-nearest ties-to-even,
  may prescribe one of those two values
  as the \textit{correct rounding} of $x$.
A floating-point implementation $\hat{f}$
  of some mathematical function $f$
  is said to be faithfully rounded
  if $\hat{f}(\hat{x})$ is a faithful rounding
  of $f(\hat{x})$ for all floating-point inputs $\hat{x}$
  in its domain,
  and likewise for correct rounding.
Correct rounding is stricter than faithful rounding:
  most real numbers have two faithful roundings
  but all have exactly one correct rounding,
  and for this reason correct rounding is required
  in many applications~\cite{why-correct-round}.

Instead of choosing a single correct rounding,
  one can also treat the two floating-point values
  $x_{\lo}$ and $x_{\hi}$ nearest $x$
  as an \emph{interval}.
Interval arithmetic then generalizes primitive operations
  to intervals instead of single floating-point values.
Instead of computing $\hat{y} = \hat{f}(\hat{x})$,
  one computes $[y_{\lo}, y_{\hi}] = \bar{f}([x_{\lo}, x_{\hi}])$
  where $y_{\lo} \le f(x) \le y_{\hi}$
  for all $x_{\lo} \le x \le x_{\hi}$.
Such an interval is called \textit{valid};
  the narrowest such interval is called \textit{tight}~\cite{ieee-1788}.
Many implementations of interval arithmetic exist
  over hardware~\cite{gaol,boost-ivals}
  and arbitrary-precision~\cite{mpfi,moore-ivals}
  floating-point;
  \toolname uses an interval arithmetic library
  called Rival~\cite{movability}.

The benefit of interval arithmetic
  is that its computations use floating-point arithmetic
  but bound the behavior of real numbers.
When applied to an expression,
  interval arithmetic produces an interval that contains
  not only the correctly-rounded result of that expression
  but \emph{all} results that can be achieved
  by faithfully rounding each primitive operation.
A reverse relationship is also true.
If a real number $x$ can be bounded to 
  some interval with relative error less than $2^{1-p}$,
  then rounding either endpoint inward
  produces a faithful rounding of $x$.
Note that the same is not true of correct rounding:
  even a very narrow interval $[x_1, x_2]$,
  much narrower than $2^{1-p}$,
  might still contain the rounding boundary
  where correct rounding switches from
  rounding up to rounding down.

\section{Mathematical Foundations}
\label{sec:bounds}

\toolname's precision assignments are based on
  a rigorous mathematical foundation
  that combines approaches from
  worst-case error bounds,
  error Taylor series,
  and condition numbers.
This mathematical foundation centers on
  three theorems.
The first,
  \Cref{eq:bound},
  is a slight variation on error Taylor series%
  ~\cite{fptaylor,hugo,satire},
  for bounding the error of expression evaluation.
The second, \Cref{eq:cnum},
  defines the $\intro$ and $\ampl_k$ functions,
  related to atomic condition numbers~\cite{atomu},
  which allow computing error bounds efficiently.
The third, \Cref{eq:constraints},
  derives sufficient conditions for
  a precision assignment.
We invite readers to focus on the theorem statements
  and skip the detailed proofs on their first read.

The basic formalism is standard.
We represent the expression to evaluate
  in ``flattened'' form,
  as a fixed sequence
  $\hat{z}_i = \hat{f}_i(\hat{x}_i, \hat{y}_i)$
  of floating-point operations $\hat{f}_i$
  on input and output floating-point registers
  $\hat{x}_i$, $\hat{y}_i$, and $\hat{z}_i$.
Each input register $\hat{x}_i$ and $\hat{y}_i$ is itself
  the output register $\hat{z}_j$ of some prior operation,
  or an argument to the expression with known value.
Naturally, some operations have one or even zero arguments
  (for floating-point constants like $\pi$),
  but for simplicity
  we write $\hat{f}_i(\hat{x}_i, \hat{y}_i)$
  for each operation.
Each register $\hat{x}_i$, $\hat{y}_i$, and $\hat{z}_i$
  has an ideal real-number value defined by $z_i = f(x_i, y_i)$,
  and arguments are assumed to carry no error.
Each $\hat{f}_i$ is assumed to be faithfully rounded
  to precision $p_i$,%
\footnote{See \Cref{sec:machine} for more
  on how this assumption is satisfied in practice}
  meaning
\[
  \hat{f}_i(\hat{x}_i, \hat{y}_i) = f_i(\hat{x}_i, \hat{y}_i) (1 + \varepsilon_i),
\]
  for some $|\varepsilon_i| \le 2^{1-p_i}$.%
\footnote{
Here we ignore under- and over-flow;
  these can be handled with
  movability flags~\cite{movability},
  or the slack mechanism of \Cref{sec:machine}.}
Substituting in the definitions of $\hat{x}_i$ and $\hat{y}_i$
  yields $\hat{z}_i$ as a function of many $\varepsilon$s,
  which we write $\hat{z}_i(\vec{\varepsilon})$.
The case of no error, $\hat{z}_i(\vec{0})$, is equal to $z_i$.

\subsection{Sound First-Order Error}

Using the definitions above, the relative error of
  the output register $\hat{z}_i$
  can be defined as
\[
\mathcal{E}[\hat{z}_i](\vec{\varepsilon}) = 
\left| \frac{\hat{z}_i - z_i}{z_i} \right| =
\left| \frac{\hat{z}_i(\vec{\varepsilon}) - \hat{z}_i(\vec{0})}{z_i} \right|
\]

\begin{theorem}\label{eq:bound}
The relative error of $\hat{z}_i$ is less than or equal to
\[
\bound(\hat{z}_i) =
\max_{\vec{\varepsilon^*}}
\sum_{j \le i} 2^{1-{p_j}}
\left|
\frac{1}{z_i}
\frac{\partial \hat{z}_i}{\partial \varepsilon_j} (\vec{\varepsilon^*})
\right|.
\]
\end{theorem}

\begin{proof}
The proof follows the typical derivation of error Taylor series,
  expanding $\mathcal{E}[\hat{z}_i]$
  into a Taylor series in the $\varepsilon_i$.
However, typically,
  an error Taylor series derivation uses
  a linear function plus a second-order term,
  $f(\varepsilon) = f(0) + \varepsilon f'(0)
                  + o(\varepsilon^2)$.
We instead apply the Lagrange remainder theorem (at order 1),
  $f(\varepsilon) = f(0) + \varepsilon f'(\varepsilon^*)$,
  with no second-order term.
Instead, a new variable $\varepsilon^*$ is introduced,
  and the first derivative is evaluated at that point.
The theorem then follows by using the multi-variable form
  of the Lagrange remainder theorem,
  substituting $\hat{z}_i$ for $f$,
  subtracting the $\hat{z}_i(\vec{0})$ term
  and dividing by $z_i$ on both sides,
  bounding each $\varepsilon_j$ by $2^{1-p_j}$,
  and using the triangle inequality to
  distribute the absolute value inside the $\max$ and sum.
\end{proof}

The intuition is that the new $\varepsilon^*$ variable
  captures the higher-order variation in $f'$.
In our context, the $\varepsilon^*$ variable represents
  a different rounding, but one that lies in
  the interval we will compute for $\hat{z}_i$.%
\footnote{Because $\vec{\varepsilon^*} \le \vec{\varepsilon}$.}
We use ``starred'' variables like $z^*_i$
  to represent the values of each intermediate variable
  with this hypothetical $\vec{\varepsilon^*}$ rounding error.

\subsection{Linear-Time Error Bounds}

Next, define the functions \intro and $\ampl_k$ by:
\[
\intro(\hat{z}_i) = 
\max_{\vec{\varepsilon^*}}\left|
\frac{f_i(x^*_i, y^*_i)}{f_i(x_i, y_i)}
\right|
\text{\quad and \quad}
\ampl_k(\hat{z}_i) = 
\max_{\vec{\varepsilon^*}}
\left|
  \frac{y_i (\partial_k f_i)(x^*_i, y^*_i)}{f_i(x_i, y_i)}
\right|
\]
  where, recall,
  $x^*_i$ and $y^*_i$ are $\hat{x}_i$ and $\hat{y}_i$
  evaluated at $\vec{\varepsilon^*}$.

\begin{theorem}\label{eq:cnum}
The expression $\bound(\hat{z}_i)$,
  defined in \Cref{eq:bound},
  is bounded by
\[
2^{1-p} \intro(\hat{z}) +
(1 + 2^{1-p}) \ampl_1(\hat{z}) \bound(\hat{x}) +
(1 + 2^{1-p}) \ampl_2(\hat{z}) \bound(\hat{y}).
\]
\end{theorem}

\begin{proof}
Expand the definition of $\bound(\hat{z})_i$
  from \Cref{eq:bound},
  and consider
  the $\partial \hat{z}_i / \partial \varepsilon_j$ term.
Substitute in the definition
  $\hat{z}_i = f_i(\hat{x}_i, \hat{y}_i) (1 + \varepsilon_i)$.
Then we have two cases:
  either $i = j$ or $i > j$.
Note that $i < j$ cannot occur,
  because steps in a computation can only reference
  earlier, not later, results.
If $i = j$,
  we have
\[
\frac{\partial \hat{z}_i}{\partial \varepsilon_j} =
\frac{\partial \hat{z}_i}{\partial \varepsilon_i} =
f_i(\hat{x}_i, \hat{y}_i),
\]
  because the $\hat{x}_i$ and $\hat{y}_i$ terms are
  computed earlier than, and are thus independent of,
  $\varepsilon_i$.
If, on the other hand, $i > j$,
  then $\varepsilon_i$ is independent of $\varepsilon_j$,
  so we have, by the chain rule,
\[
\frac{\partial \hat{z}_i}{\partial \varepsilon_j} =
(1 + \varepsilon_i) \left(
(\partial_1 f_i)(\hat{x}_i, \hat{y}_i)
\frac{\partial \hat{x}_i}{\partial \varepsilon_j}
+
(\partial_2 f_i)(\hat{x}_i, \hat{y}_i)
\frac{\partial \hat{y}_i}{\partial \varepsilon_j}
\right),
\]
where $\partial_1 f$ and $\partial_2 f$ refer to
  the derivatives of $f$ in its first and second arguments.
Performing some rearrangements
  and noting that $|\varepsilon_i| < 2^{1-p_i}$,
  we now have:
\begin{align*}
\overbrace{
\max_{\vec{\varepsilon^*}}
\sum_{j \le i} 2^{1-p_j}
\left|
\frac{1}{z_i}
\frac{\partial \hat{z}_i}{\partial \varepsilon_j} (\vec{\varepsilon^*})
\right|
}^{\bound(\hat{z}_i)}
&\le
2^{1-p_i}
\overbrace{
\max_{\vec{\varepsilon^*}}\left|
\frac{f_i(x^*_i, y^*_i)}{f_i(x_i, y_i)}
\right|
}^{\intro(\hat{z}_i)}
\\&+
(1 + 2^{1-p_i})
\left(
\max_{\vec{\varepsilon^*}}
\left|
  \frac{x_i (\partial_1 f_i)(x^*_i, y^*_i)}{f_i(x_i, y_i)}
\right|
\right)
\left(
  \max_{\vec{\varepsilon^*}}
  \sum_{j < i} 2^{1-p_j}
  \left|
    \frac1{x_i}
    \frac{\partial \hat{x_i}}{\partial \varepsilon_j}
    (\vec{\varepsilon^*})
  \right|
\right)
\\&+
(1 + 2^{1-p_i})
\underbrace{
\left(
\max_{\vec{\varepsilon^*}}
\left|
  \frac{y_i (\partial_2 f_i)(x^*_i, y^*_i)}{f_i(x_i, y_i)}
\right|
\right)
}_{\ampl_k(\hat{z}_i)}
\underbrace{
\left(
  \max_{\vec{\varepsilon^*}}
  \sum_{j < i} 2^{1-p_j}
  \left|
    \frac1{y_i}
    \frac{\partial \hat{y_i}}{\partial \varepsilon_j}
    (\vec{\varepsilon^*})
  \right|
\right)
}_{\bound(\hat{y}_i)}
\end{align*}
Substituting $\intro$ and $\ampl_k$ as per the braces
  yields the theorem.
\end{proof}

The intuition behind this theorem
  is that $\hat{z}_i$
  introduces approximately $2^{1-p_i}$ error
  and also amplifies error in its inputs
  by approximately $\ampl_k(\hat{z}_i)$.
The error bound for $\hat{z}_i$
  then sums these three sources of error.
Readers familiar with condition numbers
  will immediately see the relation to the
  atomic condition formula
  $e_z = \mu + \Gamma_x e_x + \Gamma_y e_y$~\cite{atomu}.
However, note that traditional atomic conditions
  only estimate the first-order error,
  while \Cref{eq:cnum} bounds the total error.%
\footnote{In fact, the name \ampl is a pun
  on both ``amplification factor'' and ''ample'' as in sound.}

\subsection{Precision Tuning}

Finally, we introduce our target precision, $t$,
  for output register $\hat{z}_i$.
\begin{theorem}\label{eq:constraints}
Suppose the precisions $p_i$ for each operation $\hat{f}_i$
  satisfy
\begin{align*}
  p_i &= \max(3, t + 2 + \lceil \log_2 \intro(\hat{z}_i) \rceil)
  \\
  1-\log_2 \bound(\hat{x}_i) &\ge t + 2 + \lceil \log_2 \ampl_1(\hat{z}_i) \rceil
  \\
  1-\log_2 \bound(\hat{y}_i) &\ge t + 2 + \lceil \log_2 \ampl_2(\hat{z}_i) \rceil.
\end{align*}
Then $\bound(\hat{z}_i) \le 2^{1-t}$ or,
  equivalently $1 - \log_2 \bound(\hat{z}_i) \ge t$.
\end{theorem}

\begin{proof}
The basic approach is to solve the bound in \Cref{eq:cnum}
  for the $p_j$ terms.
Formally, we seek $p_j$ such that
  $\bound(\hat{z}_i) \le 2^{1-t}$
  for some target precision $t$.
To do so, split the allowed $2^{1-t}$ error equally
  among the three terms on the right hand side of \Cref{eq:cnum}.
  
The first term yields
\[
2^{1-p_i} \intro(\hat{z}_i) \le \frac13 2^{1-t}
\]
Rearranging a little and noting that $\log_2 3 \le 2$,
  we have $p_i \ge t + 2 + \lceil \log_2 \intro(\hat{z}_i) \rceil$.

The other two terms are similar; consider, say, the first:
\[
(1 + 2^{1-p_i}) \ampl_1(\hat{z}_i) \bound(\hat{x}_i) \le \frac13 2^{1-t}.
\]
It can be rearranged into
  $1-\log_2 \bound(\hat{x}_i) \ge t + \log_2(3 + 3\cdot2^{1-p_i})
  + \log_2 \ampl_1(\hat{z}_i)$.
As long as $p_i \ge 3$,
  the term $\log_2 (3 + 3\cdot2^{1-p_i}) \le 2$
  so we must bound $\hat{x}_i$'s error
  to at most $t + 2 + \lceil\log_2 \ampl_1(\hat{z}_i) \rceil$ bits, yielding the theorem.
\end{proof}

Intuitively, \Cref{eq:constraints} states that
  the root operation must be evaluated at
  precision $t$ (plus a few extra bits),
  and then each argument's target precision
  is increased to accommodate
  the amplification of its error.
If the $\intro$ and $\ampl_k$ terms can be computed,
  \Cref{eq:constraints} then provides a recipe
  for computing the precision assignments $p_i$ from them.

Note that \Cref{eq:constraints} guarantees
  the \emph{soundness} of the precision assignment;
  that is, it guarantees that the precision assignment
  is \emph{high enough} to prevent re-evaluation,
  at least if $\intro$ and $\ampl_k$ terms can be computed.
This limits the \emph{number} of re-evaluations,
  but not their \emph{cost}, which requires choosing
  \emph{low enough} precisions for evaluation to be fast.
That said, in practice, we find it to assign
  relatively low precisions, and \Cref{sec:opt} describes
  additional practical optimizations in \toolname
  that reduce the precision needed,
  though in some cases by sacrificing or limiting
  the soundness property of \Cref{eq:constraints}.
\Cref{sec:eval} evaluates the quality
  of \toolname's precision assignments empirically.

\section{Exponent Tricks}
\label{sec:tuning}
Computing $\intro$ and $\ampl_k$ terms
  requires handling $\vec{\varepsilon^*}$.
Recall that $0 \le \vec{\varepsilon^*} \le \vec{\varepsilon}$.
Any interval containing
  $\hat{z}_i(\vec{0})$ and $\hat{z}_i(\vec{\varepsilon})$
  thus also contains  $\hat{z}_i(\vec{\varepsilon^*})$.
This is convenient, because Ziv's method naturally computes
  intervals $\bar{z}_i$ for each output register $\hat{z}_i$,
  and those intervals contain
  $\hat{z}_i(\vec{0})$ and $\hat{z}_i(\vec{\varepsilon})$,
  and so therefore also contain $\hat{z}_i(\vec{\varepsilon^*})$.

\begin{table}[p]
    \centering
    \renewcommand{\arraystretch}{1.3}
    \begin{tabular*}{\columnwidth}{c|l|l}
        Operation ($f$) & Amplification factor ($\ampl_k(f)$) & Exponent tricks bound for $\lceil \log_2 \ampl_k \rceil$\\\hline
        %%%%%%%%%%%%%%%%%%%%%%%%%%%%%%%%%
        
        $x+y$ & 
        \makecell[l]{
                    $k = 1: |x / (x + y) |$ \\
                    $k = 2: |y / (x + y) |$} &
        \makecell[l]{
          $k = 1 : \maxlog(\bar{x}) - \minlog(\bar{z})$ \\
          $k = 2 : \maxlog(\bar{y}) - \minlog(\bar{z})$
        }\\\hline
        %%%%%%%%%%%%%%%%%%%%%%%%%%%%%%%%%
        
        $x-y$ & 
        \makecell[l]{
                $k = 1: |x / (x - y) |$ \\
                $k = 2: |y / (x - y) |$} &
        \makecell[l]{
          $k = 1 : \maxlog(\bar{x}) - \minlog(\bar{z})$ \\
          $k = 2 : \maxlog(\bar{y}) - \minlog(\bar{z})$
        } \\\hline
        %%%%%%%%%%%%%%%%%%%%%%%%%%%%%%%%% 
        
        $x \times y$ &
        \makecell[l]{
                $k = 1: |y^* / y|$ \\
                $k = 2: |x^* / x|$} &
        \makecell[l]{
          $k = 1 : \logspan(\bar{y})$ \\
          $k = 2 : \logspan(\bar{x})$
        }\\\hline
        %%%%%%%%%%%%%%%%%%%%%%%%%%%%%%%%%
        
        $x \div y$ &
        \makecell[l]{
                $k = 1: |y / y^*|$ \\
                $k = 2: \left|\frac{y x^*}{{y^*}^2} / \frac{x}{y}\right|$} &
        \makecell[l]{
          $k = 1 : \logspan(\bar{y})$ \\
          $k = 2 : \logspan(\bar{x}) + 2  \logspan(\bar{y})$
        }\\\hline
        %%%%%%%%%%%%%%%%%%%%%%%%%%%%%%%%%
        
        $\sqrt{x}$ &
        $|1/2 \sqrt{x/x^*}|$ &
        $\frac12 \logspan(\bar{x}) - 1$
        \\\hline
        %%%%%%%%%%%%%%%%%%%%%%%%%%%%%%%%%

        $\sqrt[3]{x}$ &
        $| 1/3 (x/x^*)^{2/3}|$ &
        $\frac23 \logspan(\bar{x}) - 1$
        \\\hline
        %%%%%%%%%%%%%%%%%%%%%%%%%%%%%%%%% 

        $\log(x)$ & 
        $\left|(x/x^*)/\log(x)\right|$ & 
        $\logspan(\bar{x}) - \minlog(\bar{z})$ 
        \\\hline
        %%%%%%%%%%%%%%%%%%%%%%%%%%%%%%%%%
        
        $e^x$ & 
        $|x e^{x^*} / e^x|$ &
        $\maxlog(\bar{x}) + \logspan(\bar{z})$ 
        \\\hline
        %%%%%%%%%%%%%%%%%%%%%%%%%%%%%%%%% 

        $x^y$, $k = 1$ & 
        $\left|x y^* (x^*)^{y^*-1} / x^y \right|$ & 
        $\maxlog(\bar{y}) + \logspan(\bar{x}) + \logspan(\bar{z})$\\\hline
        %%%%%%%%%%%%%%%%%%%%%%%%%%%%%%%%%
        
    \end{tabular*}
    \smallskip
    \caption{Bounds on amplification factors
      using only the exponents of already-computed intervals
      $\bar{x}$, $\bar{y}$, and $\bar{z}$.
    The formulas in this table do not use
      any approximations.
    }
    \label{tab:exp-exact}

    \renewcommand{\arraystretch}{1.3}
    \begin{tabular*}{\columnwidth}{c|l}
        Operator ($f$) & Approximation used \& exponent tricks bound for $\lceil \log_2 \ampl_k \rceil$\\\hline
        
        $x^y$, $k = 2$ & 
        $\log_2 |x| \le |x| - \frac12$ \\\hline
        & $\maxlog(\bar{y}) + \max(|\minlog(\bar{x})|, |\maxlog(\bar{x})|) - 1 + \logspan(\bar{z})$
        \\\hline
        %%%%%%%%%%%%%%%%%%%%%%%%%%%%%%%%%
        
        $\cos(x)$ & 
        $|\sin(x)| \le \min(|x|, 1)$  \\\hline
        & $\maxlog(\bar{x}) - \minlog(\bar{z}) + \min(\maxlog(\bar{x}),0)$
         \\\hline
        %%%%%%%%%%%%%%%%%%%%%%%%%%%%%%%%% 
        
        $\sin(x)$ & 
        $|\cos(x)| \le 1$ \\\hline& 
        $\maxlog(\bar{x}) - \minlog(\bar{z})$ 
        \\\hline
        %%%%%%%%%%%%%%%%%%%%%%%%%%%%%%%%% 
        
        $\tan(x)$ & 
        $\max(|\sin(x)|, |\cos(x)|) \ge \frac12$ \\\hline&  
        $\maxlog(\bar{x}) + \max(|\minlog(\bar{z})|, |\maxlog(\bar{z})|) + \logspan(\bar{z}) + 1$ 
        \\\hline
        %%%%%%%%%%%%%%%%%%%%%%%%%%%%%%%%% 

        $\cosh(x)$ & 
        $|\tanh(x)| \le \min(|x|, 1)$ \\\hline& 
        $\maxlog(\bar{x}) + \min(\maxlog(\bar{x}), 0)$
        \\\hline
        %%%%%%%%%%%%%%%%%%%%%%%%%%%%%%%%%
        
        $\sinh(x)$ & 
        $|\tanh(x)| \le \min(|x|, 1)$ \\\hline& 
        $\maxlog(\bar{x}) - \min(\minlog(\bar{x}), 0)$
        \\\hline
        %%%%%%%%%%%%%%%%%%%%%%%%%%%%%%%%%

        $\tanh(x)$ & 
        $|1/\cosh^2(x)| \le 1$ \\\hline& 
        $\maxlog(\bar{x}) - \minlog(\bar{z})$
        \\\hline
        %%%%%%%%%%%%%%%%%%%%%%%%%%%%%%%%%

        $\operatorname{atan}(x)$ & 
        $1/(1 + x^2) \le 1$ \\\hline& 
        $\maxlog(x) - \minlog(\bar{z})$ 
        \\\hline
        %%%%%%%%%%%%%%%%%%%%%%%%%%%%%%%%%

        $\operatorname{atan2}(y, x)$ &
        $x^2 + y^2 \ge 2\times\min(|x|, |y|)^2, \log_2(|x^2 + y^2|) \ge 1+ 2\times\log_2\min(|x|, |y|)$ \\\hline& 
        $\maxlog(\bar{x}) + \maxlog(\bar{y}) - 1 - 2\min(\minlog(\bar{x}),\minlog(\bar{y})) - \minlog(\bar{z})$
        \\\hline
    \end{tabular*}
    \smallskip
    \caption{
    Exponent trick bounds derived via approximation.
    Most approximations, including for
      $\tan$, $\operatorname{atan}$, and $\cosh$,
      over-approximate by at most a factor of two.
    The bounds for $\sin$ and $\cos$
      over-approximate by more,
      though on relatively rare inputs.}
    \label{tab:exp_tricks}
\end{table}

\subsection{Exponent tricks}

Consider $\intro(\hat{z}_i)$,
  which is defined as
  $| \hat{z}_i(\vec{\varepsilon^*}) / \hat{z}_i(\vec{0})|$,
  maximized over all $\vec{\varepsilon^*}$.
Per the logic above,
  both the numerator and denominator of this fraction
  are contained in $\bar{z}_i$.
Now, if $\bar{z}_i$ contains zero or infinity,
  this isn't very useful---%
  in that case, the interval only bounds
  $\intro(\hat{z}_i)$ by infinity---%
  but otherwise we have
\[
\log_2 \intro(\hat{z}_i) \le
  \lceil \log_2 \hi(|\bar{z}_i|) \rceil -
  \lfloor \log_2 \lo(|\bar{z}_i|) \rfloor \le
  \underbrace{
    \lfloor \log_2 \hi(|\bar{z}_i|) \rfloor + 1
  }_{\maxlog \bar{z}_i} -
  \underbrace{
    \lfloor \log_2 \lo(|\bar{z}_i|) \rfloor,
  }_{\minlog \bar{z}_i}
\]
  where $\lo(\bar{z})$ and $\hi(\bar{z})$
  are the low and high endpoint of $\bar{z}$.

Since the right-hand side is an integer,
  our final result, then,
  is that $\lceil \log_2 \intro(\hat{z}_i) \rceil \le
  \maxlog \bar{z}_i - \minlog \bar{z}_i$.
The term $\maxlog \bar{z}_i - \minlog \bar{z}_i$
  comes up repeatedly, so we shorten it
  to $\logspan \bar{z}_i$.
Importantly,
  $\lfloor \log_2 |x| \rfloor$,
  for a normal floating-point number $x$,
  is just $x$'s exponent in its floating-point representation,
  which in MPFR is stored as a signed machine integer.%
\footnote{Technically, in MPFR,
  the exponent is offset by 1 from this definition,
  which \toolname corrects for.}
This means that $\maxlog \bar{z}_i$
  and $\minlog \bar{z}_i$
  can be ``computed'' by just reading
  already-computed machine integers from memory,
  and further operations on them just use
  ordinary machine integer arithmetic,
  which is very fast.%
\footnote{Assuming, again,
  that $\bar{z}_i$ contains neither zero nor infinity.}
We call this the ``exponent trick'':
  we can soundly approximate $\intro$ using only
  machine integer operations on
  the already-computed intervals $\bar{x}$.

Moreover, similar exponent tricks exist for
  the $\ampl_k$ factors of the arithmetic operations
  and many common elementary functions.
For addition and subtraction,
  $\log_2 \ampl_1(\hat{x} + \hat{y})$ 
  expands to $\log_2 | x / z |$,
  which is included in $\log_2 | \bar{x} / \bar{z} |$
  and thus bounded by $\maxlog \bar{x} - \minlog \bar{z}$.
For multiplication, an extra \logspan factor appears:
  $\log_2 \ampl_1(\hat{x} \cdot \hat{y})$
  is equal to $\log_2 | x y^* / x y |$;
  the $x$ terms cancel leaving a bound of $\logspan \bar{y}$.
Division, square, and cube roots  likewise
  leave $\logspan$ terms,
  shown in Table~\ref{tab:exp-exact}.

Exponents, logarithms, and the power function
  are likewise simple.
For the power function,
\[
\ampl_1(\pow(\hat{x}, \hat{y})) =
\left|
\frac{x y^* \pow(x^*, y^* - 1)}{\pow(x, y)} \right| =
\left|
\frac{x y^* \pow(x^*, y^*)}{x^* \pow(x, y)} \right| =
\left|
y^* \frac{x}{x^*} \frac{z^*}{z} \right|,
\]
  which results in the exponent tricks formula
  $\maxlog \bar{y} + \logspan \bar{x} + \logspan \bar{z}$.
Exponents and logarithms
  use similar algebraic rearrangement,
  resulting in the formulas of Tables~\ref{tab:exp-exact}.
All such formulas use
  only integer operations on \maxlog and \minlog
  of $\bar{x}$, $\bar{y}$, and $\bar{z}$
  and are thus very fast to compute.

\subsection{Approximations}

However, for some operators $f$,
  expanding $\ampl_k$ results in a $(\partial_k f)(x, y)$ term
  that must be over-approximated to construct
  exponent trick formulas for $\ampl_k(f)$.
\Cref{tab:exp_tricks} shows 
  all of the approximations used 
  and the exponent trick formulas they result in;
  the rest of this section merely contains
  detailed derivations for each row of that table.

\paragraph{Power}
The $\ampl_2$ factor for the power function $x^y$
  has uses a $\log(x^*)$ term as:
  \[
\ampl_2(\pow(\hat{x}, \hat{y})) =
\left|
\frac{y\pow(x^*, y^*)\log(x^*)}{\pow(x, y)} \right|
=
\left|
\frac{y z^* \log(x^*)}{z} 
\right|.
\]
This means that we must bound $\log_2 |\log(x)|$
  in terms of $\log_2 |x|$.
To do so, we first rewrite
\[
\log_2 |\log(x)| = \log_2 \left|\frac{\log_2(x)}{\log_2(e)}\right| = \log_2 |\log_2(x)| - \log_2 \log_2 e
\le \log_2 |\log_2(x)| - \frac12.
\]
Set $s = \log_2(x)$; since
  $\log_2 |s| \le |s| - \frac12$,
  we have $\log_2 |s| - \frac12 \le |s| - 1$.
This ultimately means that
  $\log_2 |\log(x^*)|$ is bounded above by
  $|\log_2 |\bar{x}|| - 1$.
That is just
  $\max(|\maxlog(\bar{x})|, |\minlog(\bar{x})|) - 1$,
  which results in the final
  amplification factor as:
\[
\lceil \log_2 \ampl_2(\pow(\hat{x}, \hat{y}))\rceil \le \maxlog(\bar{y}) + \max(|\maxlog(\bar{x})|, |\minlog(\bar{x})|) - 1 + \logspan(\bar{z})
\]

\paragraph{Sine and Cosine}
The amplification factor for $\sin(x)$ involves $\cos(x^*)$ as:
\[
\ampl_1(\sin(\hat{x})) = 
\left| \frac{x \cos(x^*)}{\sin(x)} \right| = 
\left| \frac{x \cos(x^*)}{z} \right| 
.\]
Because $|\cos(x)| \le 1$, we have $\log_2 |\cos(x^*)| \le 0$;
  this allows us to show that
\[
  \lceil \log_2 \ampl_1(\sin(\hat{x})) \rceil \le \maxlog(\bar{x}) - \minlog(\bar{z}).
\]

Likewise, for cosine we must bound $\sin(x^*)$,
  which shows up as:
  \[
\ampl_1(\cos(\hat{x})) = 
\left| \frac{x \sin(x^*)}{\cos(x)} \right| = 
\left| \frac{x \sin(x^*)}{z} \right| 
;\]
  one simple approximation is
  $|\sin(x)| \le \min(|x|, 1)$,
  meaning $\log_2 |\sin(x^*)| \le \min(\maxlog(\bar{x}), 0)$.
This produces the approximation
\[
  \lceil \log_2 \ampl_1(\cos(\hat x)) \rceil 
  \le \maxlog(\bar{x}) - \minlog(\bar{z}) + \min(\maxlog(\bar{x}), 0).
\]

\paragraph{Tangent}
The amplification factor for $\tan(\hat{x})$
  is:
\[
\ampl_1(\tan(\hat{x})) = 
\left| \frac{x}{\tan(x) \cos^2(x^*)} \right| = 
\left| \frac {x}{z \cos^2(x^*)}\right|
.\]
To bound its logarithm,
  we bound $\cos(x)$ in terms of $\tan(x)$.
Since $\sin(x)^2 + \cos(x)^2 = 1$,
  $\sqrt{1/2} \le \max(|\sin(x)|, |\cos(x)|) \le 1$.
Divide both sides by $|\cos(x)|$
  and take the logarithm;
  this yields:
\[
  -\frac12 - \log_2 |\cos(x)| \le \max(\log_2 |\tan(x)|, 0) \le
  - \log_2 |\cos(x)|,
\]
  or, in other words,
\[
  - 2 \log_2 |\cos(x)| \in
  2\max(\log_2(|\tan(x)|), 0) + [0, 1],
\]
  leaving
\begin{align*}
\lceil \log_2\ampl_1(\tan(\hat{x})) \rceil
&\le
\maxlog (\bar{x}) - \minlog (\bar{z}) + 2 \max(\maxlog (\bar{z}), 0) + 1 \\
&= \maxlog (\bar{x}) - \minlog (\bar{z}) + \maxlog (\bar{z}) + \max(\maxlog (\bar{z}), -\maxlog (\bar{z})) + 1 \\
&= \maxlog (\bar{x}) + |\maxlog (\bar{z})| + \logspan (\bar{z}) + 1,
\end{align*}
  which now only uses already-computed intervals
  and is at most one bit from being tight.

\paragraph{Arctangent}
The amplification factor for the arctangent
  contains a $1 / ({(x^*)}^2 + 1)$ term as:
  \[
\ampl_1(\atan(\hat{x})) = 
\left| \frac{x}{(1+ {(x^*)}^2) \atan(x)} \right| =
\left| \frac{x}{(1+{(x^*)}^2) z} \right|
,\]
  
  which we bound by using the fact, that
  $1/(1 + x^2) \le 1$, which ultimately means that
  $\log_2|1/(1 + x^2)| \le 0$ and can be pruned away.
This results in the following approximation:
\[
\lceil \log_2\ampl_1(\atan(\hat{x})) \rceil \le
\maxlog(x) - \minlog(\bar{z}).
\]

% The amplification factor for the arctangent
%   contains a $1 / ({(x^*)}^2 + 1)$ term as:
%   \[
% \ampl_1(\atan(\hat{x})) = 
% \left| \frac{x}{(1+ {(x^*)}^2) \atan(x)} \right| =
% \left| \frac{x}{(1+{(x^*)}^2) z} \right|
% ,\]
  
%   which we must bound in terms of $x$.
% Some rearrangement yields
% \[
% \log_2 \left| \frac{x}{{(x^*)}^2 + 1} \right|
% = - \log_2 \left| \frac{x^*}{x}x^* + \frac1x \right|.
% \]
% Now consider $x \ge 1$;
%   for these inputs, we clearly have
% \begin{align*}
%   -\log_2 \left| \frac{x^*}{x}x^* + \frac{1}{x} \right| &\le - \log_2 \left| \frac{x^*}{x}x^*\right| = - \log_2 \left|\frac{x^*}{x}\right| - \log_2 |x^*| = -\log_2 \left|\frac{x^*}{x}\right| - \left|\log_2 |x^*|\right| \\
%   &\le \todo{\logspan(\bar{x})} - \min(|\minlog(\bar{x})|, |\maxlog(\bar{x})|)
% \end{align*}
% Since the left- and right-hand sides
%   are both invariant under both
%   $x \mapsto 1/x$ and $x \mapsto -x$,
%   this inequality is then true for all $x$,
%   which result in the following approximation:
% \[
% \lceil \log_2\ampl_1(\atan(\hat{x})) \rceil \le
% - \min(|\minlog(\bar{x})|, |\maxlog(\bar{x})|) - \minlog(\bar{z}) \todo{+ \logspan(\bar{x})}
% \]
% Moreover, this approximation is off by at most one bit.

\paragraph{Hyperbolic Functions}

The amplification factors for $\sinh$ and $\cosh$
  both use $\tanh(x)$, as:
\begin{align*}
\ampl_1(\sinh(\hat{x})) = \left| \frac{x\cosh(x^*)}{\sinh(x)} \right|,\\
\ampl_1(\cosh(\hat{x})) = \left|\frac{x\sinh(x^*)}{\cosh(x)} \right|
.\end{align*}
Because $|\tanh(x)| \le \min(|x|, 1)$,
  we have
  $\lceil \log_2 |\tanh(x)| \rceil \le \min(\maxlog(\bar{x}), 0)$,
  which gives the following approximations:
\begin{align*}
&\lceil \log_2\ampl_1(\sinh(\hat{x})) \rceil \le \maxlog(\bar{x})  - \min(\minlog(\bar{x}), 0),\\
&\lceil \log_2\ampl_1(\cosh(\hat{x})) \rceil \le \maxlog(\bar{x}) + \min(\maxlog(\bar{x}), 0),
\end{align*}
  the approximations being off by at most one bit.

The amplification factor for $\tanh(x)$:
\[
\ampl_1(\tanh(\hat{x})) = \left| \frac{x}{\tanh(x)\cosh^2(x^*)} \right| = 
\left| \frac{x}{z\cosh^2(x^*)} \right|
,\]
  requires approximation of $\left|\frac{1}{\cosh^2(x^*)}\right|$.
We use the fact that $\cosh^2(x) \ge 1$, which essentially 
  means that $-\log_2|\cosh^2(x)| \le 0$ and can 
  be pruned away.
This results in the following approximation:
\[
\lceil \log_2\ampl_1(\tanh(\hat{x})) \rceil \le \maxlog(\bar{x}) - \minlog(\bar{z})
.\]

\section{Precision Assignment}
\label{sec:machine}
We implement all of the exponent tricks of
  \Cref{sec:tuning} in a tool called \toolname.
\toolname provides a simple interface:
  the user first \emph{compiles} an expression,
  which may have free variables,
  and then \emph{applies} the result
  to specific values for those free variables.
The compilation step converts the input expression
  into a simple register machine,
  and the application step alternates
  between evaluating the compiled instructions
  and tuning their precisions,
  repeating until the output interval is accurate enough.
Compilation is parameterized by
  a user-configurable target format and precision,
  as well as a maximum precision bound $P_{\max}$
  to control runtime on pathological inputs.

\subsection{Compilation}

\toolname supports expressions containing the constants
  $\pi$ and $e$, the standard arithmetic operators
  (plus, minus, times, divide),
  and every function specified in the POSIX standard
  for \textsf{math.h},
  including square and cube roots;
  exponents, powers, and logarithms;
  trigonometric, arc-trigonometric,
  hyperbolic and arc-hyperbolic functions;
  the error and gamma functions;
  modulus and remainder operators;
  rounding functions;
  and miscellaneous others like $\min$ and $\max$.
It also supports comparison functions,
  boolean connectives, and \textsf{if} statements.
To compile an expression,
  \toolname converts the expression to
  a sequence of instructions in three-argument form,
  where each instruction is an interval operator
  provided by the Rival library~\cite{movability}.
Common subexpression elimination is also performed.

\subsection{Application}\label{subsec:application}
When a register machine is applied to an input,
  \toolname initializes each instruction $z_i$'s
  precision $P[z_i]$
  by assuming each $\intro$ and $\ampl_k$ term is zero.
It then evaluates every instruction, in order,
  at this initial precision.
Once this first, untuned evaluation is complete,
  \toolname enters its tuning loop.

In each loop iteration,
  \toolname first checks whether the final register,
  corresponding to the expression's value,
  is sufficiently accurate;
  if so, it returns this value to the user.
Otherwise, \toolname must then derive
  new precision assignments for each operation.
The precision assignment algorithm is, basically,
  \Cref{eq:constraints} converted to code,
  filling $P[z_i]$ in reverse,
  roughly like so:
\begin{align*}
&P[z_n] = t \\
&\mathbf{for}\:i \in [n, \dotsc, 1]\:\{ \\
&\quad P[x_i] := \max(P[x_i], P[z_i] + 2 + \lceil \log_2 \ampl_1(\hat{z}_i) \rceil) \\
&\quad P[y_i] := \max(P[y_i], P[z_i] + 2 + \lceil \log_2 \ampl_2(\hat{z}_i) \rceil) \\
&\quad P[z_i] := \max(3, P[z_i] + 2 + \lceil \log_2 \intro(\hat{z}_i) \rceil) \\
&\}
\end{align*}
The loop iterates in reverse to propagate
  target precisions from output to input registers,
  just as in \Cref{eq:constraints}.
Before each iteration of this loop,
  $P[z_i]$ contains the target precision of instruction $i$.
The $P[x_i]$ and $P[y_i]$ lines correspond to
  the last two lines of \Cref{eq:constraints},
  updating the target precisions of the input registers,
  while the $P[z_i]$ line corresponds to
  the first line of \Cref{eq:constraints},
  setting $z_i$'s final computed precision $p_i$.
Once the loop is complete,
  each entry $P[z_i]$ holds the new precision assignment $p_i$.

\subsection{The Slack Mechanism}

The $\intro$ and $\ampl$ formulas
  of Tables~\ref{tab:exp-exact} and~\ref{tab:exp_tricks}
  do not work when $\bar{x}_i$, $\bar{y}_i$, or $\bar{z}_i$
  contain zero or infinity:
  the $\maxlog$s and $\minlog$s become infinite.
Intuitively, what's happening is that, in this case,
  the relative error represented by the interval is infinite,
  so the precision assignment algorithm does not know
  how many bits will suffice
  to produce a sufficiently narrow interval.
Another mechanism is needed.

\toolname replaces the infinite value with a \textit{guess}:
  $\slack[n]$, where $n$ is the tuning iteration number.
For example, the interval $\bar{z} = [-2^{-93}, 2^{-108}]$ 
  crosses zero, so $\minlog(\bar{z}) \in (-\infty, -108]$;
  the slack mechanism guesses that it is $-108 - \slack[n]$.
Similarly, if $\bar{z} = [2^{73}, \infty]$,
  the slack mechanism guesses $\maxlog(\bar{z}) = 73 + \slack[n]$.

\toolname sets $\slack[n]$ to $t 2^{(n-1)}$: 
  the target precision initially, doubled at each iteration.
This roughly matches
  the standard precision-doubling implementation
  of Ziv's strategy, in that it involves
  guessing a precision, testing it,
  and doubling the precision if the guess fails.
However, it's critical to understand that,
  while Ziv's strategy involves \emph{expression-wide} guesses,
  \toolname's guesses are \emph{per-primitive}.
It's common for \toolname to need guesses for \emph{some}
  of the primitives in an expression,
  but to analytically derive error bounds for others.
In fact, since the slack mechanism applies occasionally,
  it can be set more aggressively, 
  without losing too much in runtime.%
\footnote{We don't do so in this paper,
  to maintain the fairness of the evaluation,
  but have found that increasing $\slack[n]$
  by about $5$--$10\times$ slightly reduces evaluation time.}

\subsection{Correct Rounding}

\toolname aims to evaluate expressions
  to their correctly-rounded double precision values.
This can be customized by the user
  by supplying a \textit{discretization}:
  a target precision $t$,
  a conversion from MPFR values to some type $T$,
  and a distance function on pairs of $T$s.
The distance function decides whether an interval's endpoints
  are equal, unequal, or \textit{neighbors} (see below).
By supplying a different discretization,
  users can select a different rounding mode (like faithful rounding)
  or number format (such as IEEE~754 decimal floating-point).

Correct rounding introduces a tricky edge case.
\toolname's precision assignments guarantee
  that the computed interval will be accurate enough
  to faithfully rounding the result.
But even an arbitrarily narrow interval
  can still straddle a rounding boundary,
  leaving the correctly-rounded result undetermined.
This occurs when the output interval's endpoints
  round to successive double-precision values
  (or, if the user supplied a discretization,
  when its distance function returns the ``neighbor'' value).
\toolname detects this and adds $\slack[n]$ 
  to the target precision $t$ before tuning,
  in effect demanding more precision to resolve
  which side of the rounding boundary the result is on.

\section{Implementing \toolname}
\label{sec:opt}

\toolname is implemented in Rust with an eye toward maximum performance.
Intervals use a port of the Rival library~\cite{movability} to Rust,
  using pre-allocated MPFR endpoints.
\toolname implements the architecture of \Cref{sec:machine}
  with several additional optimizations
  described below.

\paragraph{Initial precisions}
To make initial evaluation as fast as possible,
  the initial precision assignment,
  which is the same for all inputs, is pre-computed
  in a separate array, avoiding the need to
  clear the precision array between inputs.
Output registers are also pre-allocated,
  as are the various bit arrays required by other optimizations.

\paragraph{Repeat operations}

It often happens that an operation is assigned
  the same precision for multiple tuning iterations.
For example, one subtree of the expression
  might trigger the slack mechanism
  while another is assigned a low precision analytically.
Reval thus sets a ``repeat bit'' for instructions
  whose precision and arguments are
  the same (or lower) than in the previous iteration;
  those instructions are skipped in later evaluations.
Note that this ``repeat optimization'' is only possible
  due to \toolname's non-uniform precisions:
  if precisions were uniform, every instruction would
  increase in precision on every iteration.

\paragraph{Constant sub-expressions}

Expressions often contain constant subexpressions
  like $\pi/2$.
\toolname detects these subexpressions during compilation
  and sets their repeat bits even during
  the initial, un-tuned evaluation
  (by modifying the pre-computed
  initial precision and repeat arrays).
The constant subexpressions are then skipped
  even during the initial evaluation.
Precision tuning is still applied to these subexpressions,
  in case they need to be recomputed at higher precision,
  at which point the higher precision is available
  for all later evaluations.
While this optimization is possible without \toolname,
  its implementation relies on repeat bits.

\paragraph{Useless intervals}

The Rival library used by \toolname
  supports ``movability flags'', which detect
  when recomputing an interval at higher precision
  will not change the interval endpoints,
  typically due to limited exponent width~\cite{movability}.
We call such intervals ``exact'', noting that
  these need not be point intervals
  (see \citet{movability} for details).
Inputs that cause a domain error,
  like $-1$ to $\sqrt{x}$,
  set ``error flags'' in the Rival library
  and are also typically exact.
To leverage these flags,
  \toolname has ``useful bits'' for each register;
  the output register is useful,
  registers with exact intervals are not,
  and arguments to useful instructions are useful.
Useless instructions have their repeat bits set
  so that they are not re-computed.
  
\paragraph{Rare operations}

The Rival interval library supports operators
  like $\operatorname{erfc}$
  that have no exponent trick formula in \Cref{tab:exp_tricks}.
\toolname just uses $\slack[n]$ for their $\ampl$ factors.
That's no better than standard precision-double algorithms,
  but these operators are rare
  so basic support seems acceptable.
Also, some operators like $\operatorname{asin}$
  use the slack mechanism for certain inputs.
For example, for $\operatorname{asin}$'s $\ampl_1$,
  \toolname uses
  $\maxlog(\bar{x}) - \minlog(\bar{z}) + 1$
  when $\maxlog(\bar{z}) \le 0$,
  but $\slack[n]$ otherwise,
  as no better exponent trick could be found.
Similarly, for the floor operator, which has no derivative,
  \toolname uses $\maxlog(\bar{x})$
  if $\bar{x}$ contains more than one integer,
  $-\infty$ if it contains zero integers,
  and $\slack[n]$ if it contains exactly one integer (a possible double rounding).

\paragraph{Loose intervals}

\Cref{eq:constraints} assumes the interval library
  produces maximally \emph{tight}~\cite{ieee-1788} intervals.
In practice, the underlying Rival interval library
  only achieves tightness up to a small multiple of~$2^{-p_i}$.
To safely apply our tuning algorithm with such libraries,
  we increase the precision of each operation by a few bits
  to provide a margin of error.
For example, in our tests,
  Rival's \textsf{fmod} has up to eight ULPs of ``looseness'';
  \toolname thus assigns up to three additional bits of precision.

\paragraph{\logspan terms}

To micro-optimize,
  \toolname initially computes precision assignments
  without including $\logspan$ terms.
Because, in the final iteration,%
  \footnote{In the final iteration,
    all values are known to at least 3 bits of precision.}
  $\logspan(\bar{z}) \le \frac14$ and
  no exponent trick formula contains
  more than three $\logspan$ terms,
  the total of all \logspan terms is less then 1,
  which is unlikely to affect the final precision.
Meanwhile, in earlier iterations,
  the $\logspan$ terms may be large,
  overestimating the necessary precision,
  so not including them is beneficial.
If the $\logspan$ terms are, in fact, necessary,
  all repeat bits end up set,
  in which case \toolname re-enables $\logspan$ terms
  to ensure soundness in the next evaluation.
This extreme edge-case
  \emph{does not actually occur in our evaluation},
  but forms a hypothetical fully-sound fall-back.

\paragraph{Error distribution}

\Cref{eq:constraints} is derived by distributing
  the error of each output register across three terms,
  assigning a third of the error bound to each.
However, a \emph{non-uniform} distribution of error
  is potentially favorable.%
\footnote{We thank the reviewers for suggesting
  this line of inquiry.}
We tested assigning
  55\% of the error to one input register,
  30\% of the error to the other input register,
  and 15\% of the error to the final $\intro$ term.
The effect of this distribution is that,
  instead of using $P[z_i] + 2$ in the updates
  for $x_i$, $y_i$, and $z_i$,
  the three registers instead use
  $P[z_i] + 1$, $P[z_i] + 2$, and $P[z_i] + 3$,
  with the $P[z_i] + 1$ term used
  for the input register with the deepest computation tree.
In theory, this should result in smaller precisions
  for deeply nested terms.
However, we were unable to measure an actual improvement,
  and stuck to the $P[z_i] + 2$ increment
  for the final version of \toolname.

\paragraph{Conclusion}

\begin{algorithm}
\caption{\toolname's precision tuning algorithm \sc{BackwardPass}.}
\label{algo:tuning_algo}
\begin{algorithmic}[1]
\Require \toolname register machine $\mathit{m}$,
    iteration $k$.
\Ensure Updates precisions and repeats vector; returns true if no reevaluations are scheduled.

\State $P' \gets \mathrm{0}$
\State $P'[z_n] \gets m.t + 3$
\Comment{1. ``Loose intervals'' optimization}

\Statex
\State $straddles \gets \mathit{m.distance}(\mathit{m.convert(m.reg[z_n].lo), m.convert(m.reg[z_n].hi)) = 1}$
\State $P'[z_n] \gets P'[z_n] + (\mathit{slack}[k]\ \mathbf{if}\ straddles\ \mathbf{else}\ 0)$
\Comment{2. Rounding boundary detection}

\Statex
\State $\mathit{m.repeats} \gets \mathrm{true}$
\Comment{3. ``Useless intervals'' optimization}
\State $\mathit{m.repeats}[z_n] \gets \mathrm{false}$
\For{$i = n, \ldots, 1$}
    \State $\mathit{m.repeats}[z_i] \gets
        \mathit{m.repeats}[z_i]
        \lor \Call{IsExact}{\mathit{m.reg}[z_i]}$
    \If{$\neg \mathit{m.repeats}[z_i]$}
        \State $\mathit{m.repeats}[x_i],\,\mathit{m.repeats}[y_i]
            \gets \mathrm{false},\,\mathrm{false}$
    \EndIf
\EndFor

\Statex
\For{$i = n, \ldots, 1$}
 \Comment{4. Precision assignment algorithm}
    \If{$\neg \mathit{m.repeats}[z_i]$}
        \State $\mathit{increment_x}, \mathit{increment_y} \gets \Call{ExpTrick}{z_i, \mathit{m.reg}[x_i], \mathit{m.reg}[y_i],
        \mathit{m.reg}[z_i],
        k, m.logspan}$
        \State $P'[x_i] \gets \min(P_{\max},
            \max(P'[x_i], P'[z_i] + 2
            + \mathit{increment_x}))$
        \State $P'[y_i] \gets \min(P_{\max},
            \max(P'[y_i], P'[z_i] + 2
            + \mathit{increment_y}))$
        \State $P'[z_i] \gets \min(P_{\max},
            \max(3, P'[z_i] + 2 +(\logspan(\mathit{m.reg}[z_i])\ \mathbf{if}\ \mathit{m.logspan}\ \mathbf{else}\ 0)))$
    \EndIf
\EndFor

\Statex
\For{$i = 1, \ldots, n$}
\Comment{5. ``Repeat operations'' optimization}
    \If{$\neg \mathit{m.repeats}[z_i]$}
    \State $\mathit{m.repeats}[z_i] \gets
        (P'[z_i] \le m.P[z_i])
        \land \mathit{m.repeats}[x_i]
        \land \mathit{m.repeats}[y_i]$
    \EndIf
\EndFor

\Statex
\State $stuck \gets \bigwedge_{i=1}^{n} \mathit{m.repeats}[z_i]$
\Comment{5. Detect whether machine is stuck and retry with $\logspan$}
\If{$stuck \land \neg m.logspan$}
\State $m.logspan \gets \mathrm{true}$
\State \Return $\Call{BackwardPass}{m,k}$
\EndIf
\Statex
\State $\mathit{m.P} \gets \mathit{P'}$
\State \Return $stuck$
\end{algorithmic}
\end{algorithm}

\begin{algorithm}
\caption{\toolname's evaluation loop}
\label{algo:evaluation_loop}
\begin{algorithmic}[1]
\Require \toolname register machine $\mathit{m}$, input point $args$ as a length-$N$ vector.
\Ensure Returns the correctly-rounded output,
    or ``$\mathrm{Unevaluable}$''.
\State $\mathit{m.repeats} \gets \mathrm{false}$, $\mathit{m.logspan} \gets \mathrm{false}$
\Comment{1. Machine initialization}
\State $\mathit{m.P} \gets \mathit{m.P_{init}}$
\Comment{1.1 ``Initial precisions'' optimization}
\State $m.reg[0:N] \gets args[0:N]$
\Comment{1.2 Storing input into machine}

\Statex
\For{$k = 1, 2, 3, \ldots$}
\Comment{2. Evaluation loop}
\For{$i = 1, \ldots, n$}
\Comment{2.1 Evaluating instructions}
 \If{$\neg \mathit{m.repeats}[z_i]$}
\State $m.reg[z_i] \gets
    \bar{f}_i\bigl(m.reg[x_i],m.reg[y_i]\bigr)$
    \textbf{at precision} $m.P[z_i]$\EndIf
\EndFor
\Statex
\If{$\mathit{m.distance}(\mathit{m.convert(m.reg[z_n].lo), m.convert(m.reg[z_n].hi)) = 0}$}
\State \Return $m.convert(m.reg[z_n].lo)$
\Comment{2.2 Correctly-rounded result is obtained}
\EndIf
\Statex

\State $\mathit{stuck} \gets \Call{BackwardPass}{m,k}$
\Comment{2.3 Precision tuning}

\If{$stuck$}
\State \Return $\mathrm{Unevaluable}$
\Comment{2.4 Exit if machine is stuck}
\EndIf
\EndFor
\end{algorithmic}
\end{algorithm}

\Cref{algo:tuning_algo} contains combined pseudo-code for 
  precision-tuning algorithm,
  integrating most of the optimizations in this section,
  while \Cref{algo:evaluation_loop} contains
  combined pseudo-code for  \toolname's evaluation loop.
Both algorithms operate on
  a compiled ``machine'' structure $m$,
  which stores fields such as the precision vector $P[z_i]$,
  registers $reg[z_i]$, repeats vector $repeats[z_i]$,
  the user-specific precision limit $P_{\max}$,
  the precomputed initial precision $P_{\mathit{init}}$
  the logspan flag $logspan$, the target precision $t$,
  and the discretization's $convert$ and $distance$ functions.

Note that, with the various optimizations in this section,
  \Cref{eq:constraints} no longer applies
  to the typical evaluation.
Strictly speaking, soundness is only guaranteed
  if none of the \textsc{ExpTrick} invocations use $\slack[n]$
  and if the $logspan$ flag is set.
However, \Cref{eq:constraints} still defines
  the overall structure of precision assignment,
  correct evaluations are still guaranteed
  by the distance check in \Cref{algo:evaluation_loop},
  and \Cref{eq:constraints} still ensures termination
  if $\slack$ is not used, after at most
  one additional evaluation (to set the $logspan$ flag).

\section{Evaluation}
\label{sec:eval}
We investigate the impact of \toolname's precision tuning algorithm
through four research questions:
\begin{enumerate}
\item[RQ1:] Is \toolname \emph{faster}
    than prior real expression evaluation methods?
\item[RQ2:] How close are \toolname's \emph{precision assignments} to optimal?
\item[RQ3:] How effective are \toolname's \emph{optimizations}
    at reducing primitive computations?
\item[RQ4:] Is \toolname practical for real-world applications?

% \item[RQ1] Can \toolname evaluate expressions \emph{faster}
%   than a uniform-precision baseline?
% \item[RQ2] Is this because \toolname uses lower precisions
%   than a uniform-precision baseline?
% \item[RQ3] How do \toolname's optimizations
%   contribute to performance improvements?
% \item[RQ4] Is \toolname practical for real-world applications?

\end{enumerate}
We evaluate RQ1 by comparing \toolname to
  the well-known Sollya package,
  and evaluate RQs 2 \& 3 by comparing \toolname
  to two ablations of \toolname itself:
  ``Ziv'', which uses
  a standard uniform precision-doubling algorithm
  with no additional optimizations,
  and ``Ziv+'',
  which uses the same precision assignments
  but incorporates \toolname's tuning-unrelated optimizations.
All tools are run on
  \NumTunedBenchmarks expressions
  totaling \NumTunedPoints inputs.

\subsection{Methodology}

\paragraph{Baselines}

Both Ziv and Ziv+ are variants of \toolname.
Both replace \toolname's precision assignment algorithm
  with a standard precision-doubling approach,
  and Ziv (but not Ziv+) additionally disables
  \toolname's constant subexpression
  and ``useless operation'' optimizations.
Note that the repeats optimization is disabled in both,
  since it requires non-uniform precision assignments
  to fire.
Sollya is an independently-developed numerics workbench.
Its expression evaluation algorithm mostly
  uses the standard precision-doubling approach,
  though Sollya can detect when it's ``close''
  to the correct precision and use heuristics
  to increase the precision by less than a factor of two;
  in any case, the precision is uniform.
It has a variant of
  \toolname's constant subexpression optimization
  but no analog of the ``useless operation''
  or repeats optimization.
We validate that all four tools agree on each input
  and set all four to the same initial and maximum precisions.

\paragraph{Expressions}
Our benchmark suite is
  \NumBenchmarksIncludeDiscarded mathematical expressions
  drawn from various FPCore-using applications~\cite{fpbench}.
Expressions vary from \NumBenchOpsMin to \NumBenchOpsMax~operations
  (average \NumBenchOpsAvg)
  and have \NumBenchVarsMin--\NumBenchVarsMax~variables
  (average \NumBenchVarsAvg).
Despite their small size, these operations are challenging,
  with some requiring
  tens of thousands of bits for certain inputs;
  hardware and software validation use cases
  typically focus on similarly-small expressions.
Arithmetic operations are the most common by far (80\%+)
  but the exponential, power, sine, cosine,
  logarithm, and square root operations
  are also common ($1.5\%$ each).
\NumBenchmarksDiscardedbyFunctionality expressions
  aren't supported by Sollya,
  and two more trigger a bug in it,%
\footnote{The two expressions are
  $z^z e^{-z}$ and $e^{-z} \ell^{\exp(z)}$;
  the bug was reported to and acknowledged by the developers.}
  leaving \NumBenchmarks total.

\paragraph{Inputs}
For each benchmark,
  we randomly sample \NumPointsPerBenchmark double-precision
  input points (that is, values for each variable)
  using rejection sampling,
  ignoring points with domain errors.
We pre-sample and pre-parse all inputs
  and use the same input points
  for both \toolname and the baselines.
\toolname, Ziv+ and Ziv are set to
  correctly-rounded double-precision;
  Sollya is set to faithful rounding.%
\footnote{
  Sollya mostly produces correctly-rounded results,
    but in \SollyaFaithfulCnt cases, roughly 1\%,
    it instead produces a faithful rounding.
}
Most of the remaining points
  are accurate enough at the starting precision
  and do not require precision tuning;
  we discard these points as well,
  leaving a total of \NumTunedPoints inputs
  across \NumTunedBenchmarks benchmarks.

\paragraph{Measurement}

We measure runtime for each input to each expression separately,
  using an in-process timer to minimize overhead.
A timeout of 50 milliseconds per input is applied;
  this mostly affects Sollya,
  which has a rare seemingly-infinite loop on some inputs.
Evaluation is performed on a machine with
  an Intel i7-8700 at 3.7\,GHz and 32\,GB of DDR4~memory,
  running Ubuntu~24.04.02;
  Sollya~8.0 with MPFI~1.5.3, MPFR~4.2.1, and GMP~6.3.0;
  and Rust~1.75.

\subsection{Results}

\begin{figure}
\includegraphics[width=0.8\linewidth,trim={0 .2in 0 .2in}]{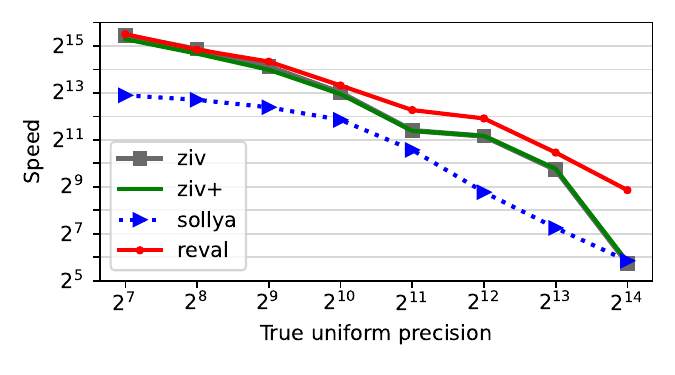}
\caption{
Evaluations per second for \toolname and Sollya,
  versus the uniform precision required for a correct evaluation,
  which measures the difficulty of expression evaluation.
\toolname is faster than Sollya at all precisions,
  $\RivalAvgSpeedupOverSollya\times$ faster on average.
Moreover, its advantage increases with difficulty,
  up to $\RivalMaxSpeedupOverSollya\times$
  on the most difficult evaluations.
The Ziv and Ziv+ ablations demonstrate that
  \toolname's advantage, especially on
  the most difficult evaluations,
  is due largely to \toolname's precision assignment algorithm.
}
\label{fig:ratio_per_precision}
\end{figure}

\paragraph{RQ1}
\Cref{fig:ratio_per_precision} shows evaluations per second
  for \toolname and Sollya,
  with evaluations aggregated by
  the uniform precision needed to evaluate the expression,
  a proxy for evaluation difficulty.
\toolname is significantly faster,
  by $\RivalAvgSpeedupOverSollya\times$ on average,
  and its advantage grows with difficulty:
  for the most difficult evaluations,
  which require precisions up to $2^{14}$,
  \toolname is $\RivalMaxSpeedupOverSollya\times$ faster than Sollya.
The scale of the speed-up shows
  that \toolname is a substantial advance
  over the state of the art.

\begin{figure}
\begin{minipage}[t]{0.48\linewidth}
\includegraphics[width=\linewidth, trim={.15in .3in .15in 0}]{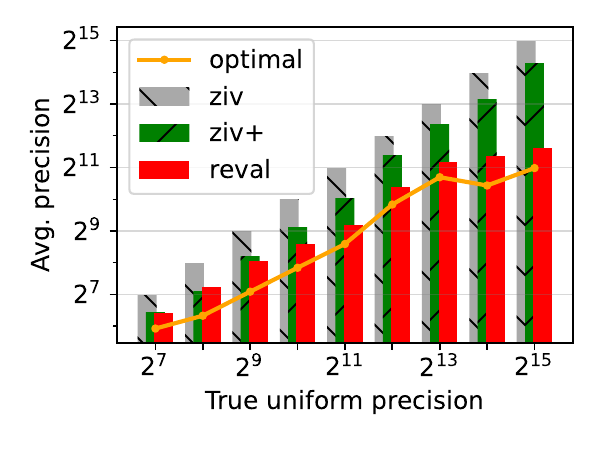}
\caption{
Averaged precisions for \toolname, Ziv+, and Ziv that are
  aggregated by the problem difficulty (high-enough uniform precision); lower is better.
\toolname precision assignments are $\RevalCloserToOptimalThanBaseline\times$ closer on average
  to the optimal ones than Ziv+'s precisions and
  $\RevalCloserToOptimalThanZiv\times$ closer than Ziv's precisions.
}
\label{fig:optimality_plot}
\end{minipage}\hfill%
\begin{minipage}[t]{0.48\linewidth}
\includegraphics[width=\linewidth, trim={.15in .3in .15in 0}]{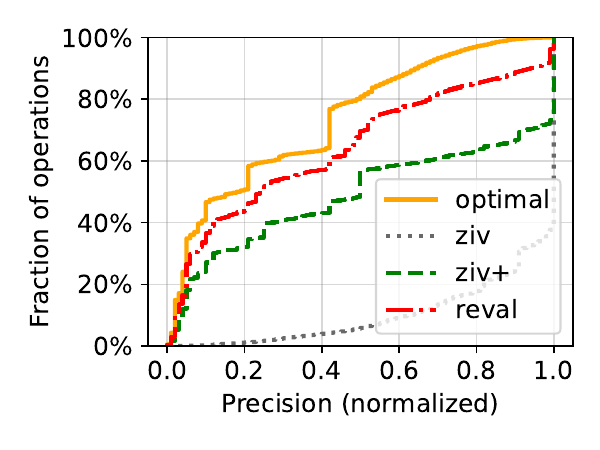}
\caption{
Distribution of final precision assignments
  per all the evaluations, normalized against a maximum recorded precision
  of \emph{all four tools} per evaluation; left is better.
\toolname manages to execute $\DensityAdvantageRevalOverBaseline\times$ more primitives
  under 20\% of a maximum recorded precision than Ziv+,
  and $\DensityAdvantageRevalOverZiv\times$ more primitives than Ziv.
\toolname also is closer to the optimal
  precision assignments than other tools
  at any distribution point.
}
\label{fig:density_cdf}
\end{minipage}
\end{figure}

The cause of this speed-up differs for different evaluations;
  to demonstrate this, \Cref{fig:ratio_per_precision} also shows
  the performance of the Ziv and Ziv+ ablations of \toolname.
For the easiest evaluations,
  Ziv and Ziv+ achieve similar performance to \toolname,
  showing that \toolname's advantage comes
  largely from choosing a faster interval arithmetic library
  (Rival 3, instead of Sollya's choice of MPFI)
  and from a well-optimized evaluator.
However, as the evaluations get more difficult,
  the performance of Ziv and Ziv+ drops,
  until for the most challenging evaluations
  Ziv and Ziv+ perform identically to Sollya.
\toolname however, not only retains but grows its advantage,
  demonstrating that its precision assignment algorithm
  is largely responsible for its good performance
  on the most challenging inputs.

% \begin{figure}
% \begin{minipage}[t]{0.4\linewidth}
% \includegraphics[width=\linewidth,trim={0 .175in 0 0}]{images/ratio_plot_precision.pdf}
% \caption{
% Evaluations per second split by 
%   true uniform precision required for a correct evaluation,
%   which essentially indicates difficulty; higher is better.
% \toolname is, on average,
%   $\RivalAvgSpeedupOverBaseline\times$ faster than Ziv+,
%   $\RivalAvgSpeedupOverZiv\times$ faster than Ziv,
%   and $\RivalAvgSpeedupOverSollya\times$ faster than Sollya.
% \toolname's advantage is
%   $\RivalMaxSpeedupOverBaseline\times$ over Ziv+,
%   $\RivalMaxSpeedupOverZiv\times$ over Ziv,
%   and $\RivalMaxSpeedupOverSollya\times$ over Sollya
%   on the most difficult points.
% }
% \label{fig:ratio_per_precision}
% \end{minipage}\hfill%
% \begin{minipage}[t]{0.58\linewidth}
% \includegraphics[width=\linewidth,trim={0 .15in 0 0}]{images/histogram_valid.pdf}
% \caption{
% The distribution of a primitive execution time
%   by precision, for \toolname, Ziv+ and Ziv;
%   lower is better.
% Ziv+ and Ziv waste a lot of time
%   re-computing the same primitives
%   at successively higher precisions,
%   leading to a lot of wasted time
%   at precisions between $2^5$ and $2^{11}$.
% The ``tuning'' column shows the run time
%   of \toolname's precision tuning algorithm,
%   plus the repeat, usefulness,
%   and constant subexpression optimizations.
% }
% \label{fig:histogram}
% \end{minipage}
% \end{figure}

\paragraph{RQ2}

To analyze how close precisions are to the optimal,
  we implemented an offline procedure that
  finds the lowest valid precision assignment
  for each expression evaluation.
This ``optimal'' assignment
  uses the lowest possible precision
  for each primitive in an expression.
We then compare this precision assignment
  to the assignments chosen by Ziv, Ziv+, and \toolname.

\Cref{fig:optimality_plot} compares
  the average precision assignment for all three tools
  to the optimal assignment.
Ziv and Ziv+ use substantially higher precisions than optimal,
  while \toolname's precision assignments closely track the optimal,
  even for the most challenging evaluations.
Note that Ziv+ improves on Ziv
  thanks to its constant subexpression
  and useless operation optimizations,
  showing that these optimizations have value,
  especially at lower difficulties.
However, \toolname's precision assignment algorithm
  substantially outperforms both Ziv and Ziv+:
  $\RevalCloserToOptimalThanBaseline\times$ closer
  to the optimal precision assignment than Ziv+,
  and $\RevalCloserToOptimalThanZiv\times$ closer than Ziv.
The gap is particularly pronounced,
  meaning that \toolname's precision assignments
  are particularly advantageous
  on the most difficult evaluations.

\Cref{fig:density_cdf} shows the cumulative distribution
  of primitive precisions within an expression.
In this plot, curves that are further up evaluate
  a larger fraction of primitives at low precision,
  so imply better precision assignments.
Ziv applies uniform precisions, %
% \footnote{The non-uniformity comes only from pre-conditions,
%   which Ziv compiles away in some cases
%   because it uses the same compiler as \toolname.}
  leading to a cumulative distribution with
  a sharp cliff near 100\% of maximum precision.
Ziv+, thanks to its
  constant subexpression and useless operation optimizations,
  closes more than half the distance to
  the optimal precision assignment.
Still, \toolname is substantially closer yet,
  closing more than half the remaining distance,
  showing again that \toolname
  achieves close-to-optimal precision assignments.
Numerically, \toolname executes
  $\DensityPercentageOfLowerPrecisionReval\%$ of primitives
  at 20\% of the highest precision per all four tools,
  while Ziv+ executes $\DensityPercentageOfLowerPrecisionBaseline\%$ of primitives,
  and Ziv performs the worst with $\DensityPercentageOfLowerPrecisionZiv\%$ of primitives.
No tool reaches the optimal
  $\DensityPercentageOfLowerPrecisionOptimal\%$ of primitives
  executed at 20\% of the highest precision,
  but nonetheless, 
  \toolname's precision assignments are clearly closest
  to the optimal ones.

\paragraph{RQ3}

\begin{figure}
\begin{minipage}[t]{0.35\linewidth}
\includegraphics[width=\linewidth, trim={.15in .3in .15in .3in}]{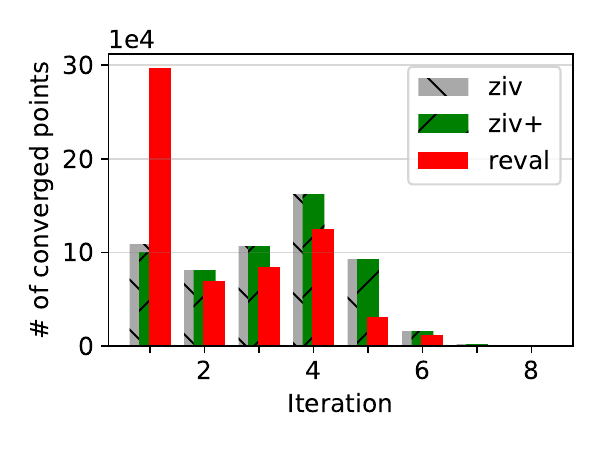}
\caption{
The number of iterations needed by \toolname, Ziv+, and Ziv;
  left is better.
\toolname requires fewer iterations than Ziv+ and Ziv
  because it derives its precision assignments
  instead of using search.
In the most of evaluations \toolname manages
  to find a high-enough precision assignment
  after only one tuning iteration.
}
\label{fig:cnt_per_iters_plot}
\end{minipage}\hfill%
\begin{minipage}[t]{0.62\linewidth}
\includegraphics[width=0.9\linewidth,trim={0 .2in 0 0}]{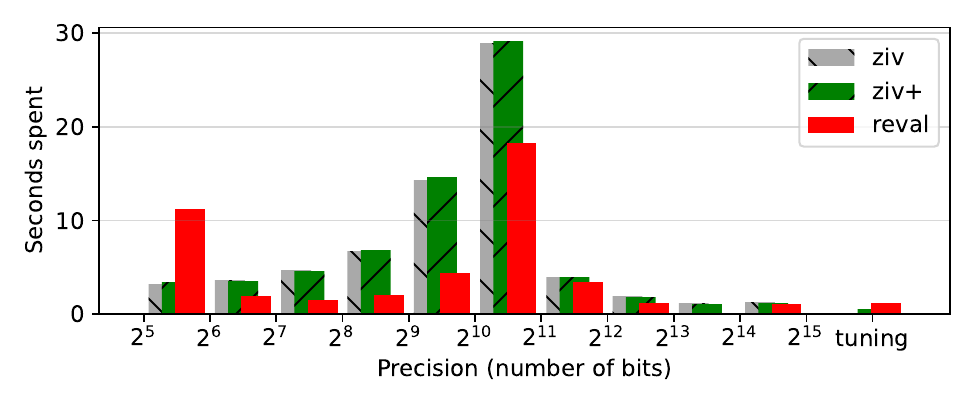}
\caption{
The distribution of a primitive execution time
  by precision, for \toolname, Ziv+ and Ziv;
  lower is better.
Ziv+ and Ziv waste a lot of time
  re-computing the same primitives
  at successively higher precisions,
  leading to a lot of wasted time
  at precisions between $2^5$ and $2^{11}$.
The ``tuning'' column shows the run time
  of \toolname's precision tuning algorithm,
  plus the repeat, usefulness,
  and constant subexpression optimizations.
}
\label{fig:histogram}
\end{minipage}
\end{figure}

\Cref{fig:cnt_per_iters_plot}
  compares the number of total tuning iterations
  required by Ziv, Ziv+, and \toolname.
\toolname performs far fewer tuning iterations, because
  instead of \emph{searching} for the correct precision
  by repeatedly trying various precisions,
  it can \emph{derive} the correct precision analytically
  in many cases.
For example, after two iterations,
  \RivalSecondIterConvergence\% of \toolname's inputs have converged while only \BaselineSecondIterConvergence\% of Ziv+'s have, 
  with Ziv performing about the same.
This leads to a radical reduction in
  the number of primitive computations needed by \toolname:
  \RivalInstrCountLessThanBaseline\% fewer than Ziv+,
  \RivalInstrCountLessThanZiv\% fewer than Ziv.

This reduction in evaluation \emph{counts}
  further contributes to a reduction in evaluation \emph{time}.
\Cref{fig:histogram} compares \toolname, Ziv and Ziv+
  on time spent computing primitives at every precision.
\toolname's precision assignments allow it to
  skip many primitive computations between $2^5$ and $2^{11}$
  by jumping straight to the correct precision assignment.
The rightmost ``tuning'' column also shows
  that precision tuning itself is very fast---%
  about \TuningTimePercentage\% of total run time.
The tuning column for Ziv+ measures its
  usefulness and constant subexpression optimizations,
  while Ziv's tuning time is zero as no tuning is performed.

\paragraph{RQ4}

% Data from https://nightly.cs.washington.edu/reports/herbie/1742604846:nightly:baseline-kernel:3191bcb3fe/timeline.html
To test whether \toolname is practical
  for real-world applications,
  we worked with the developers
  of the Herbie numerical compiler~\cite{herbie}
  to integrate \toolname into its ``sampling'' phase.
This phase evaluates a given expression
  on thousands of input points,
  which are then used as a ``golden model''
  to test programs synthesized by Herbie.
Prior to integrating \toolname,
  Herbie used the standard uniform precision-doubling approach,
  but performance was a problem:
  sampling was $40\%$ of Herbie's runtime,
  making it difficult to apply Herbie to larger programs.
Our integration effort took several months,
  and required introducing
  significant new abstractions
  around real evaluation in Herbie.
The result was a significant speedup,
% Data from https://nightly.cs.washington.edu/reports/herbie/1742600049:nightly:rival-kernel:72e83ef020/timeline.html
  decreasing sampling time from 16.7~to 12.2~minutes,
  a $\RivalHerbieAverageSpeedup\times$ speedup.
Note that this speedup is significantly smaller than
  the $\RivalAvgSpeedupOverSollya\times$ speedup of RQ1,
  because sampling in Herbie involves
  random number generation, logging, format conversion,
  and FFI overhead, and also includes inputs
  that do \emph{not} require precision tuning,
  all of which \toolname does not speed up.
Our integration work was included in a recent release,
  and public Herbie installations now use \toolname by default.

% \todo{remove} This last category demands some elaboration.
% A real-world tool like Herbie sees
%   tens of thousands of ``difficult inputs'',
%   which require more precision to evaluate
%   than Herbie's maximum precision of 10\thinspace000~bits.
% \toolname is significantly faster on these difficult inputs;
%   we captured and retested these inputs,
%   and found that \toolname exits
%   $\RivalExitTimetoBaseline\times$ faster than Ziv+
%   (and $\RivalExitTimetoSollya\times$ faster than Sollya).
% This is because \toolname can often assign
%   a very high precision after just one or two iterations,
%   often without \emph{any} very-high-precision computations.
% Ziv+ and Sollya, meanwhile,
%   must try very high precisions before giving up.
% \Cref{fig:histogram-all} replicates \Cref{fig:histogram}
%   but with these difficult inputs included:
%   Baseline now spends a huge amount of time
%   on fruitless evaluations at high precision,
%   while \toolname spends very little additional time.
% Before integrating \toolname,
%   these difficult inputs were $36\%$
%   of Herbie's total time evaluating real expressions;
%   after integrating \toolname,
%   they were only $24\%$.
% Note that these inputs are ignored in RQs 1--3;
%   \toolname exceptional performance on them
%   means its real-world impact is \emph{larger}
%   than in the apples-to-apples comparison above.

% \smallskip
% \todo{\st{The Herbie developers plan for their next release, Herbie~2.2,
%   to use \toolname for real evaluation.}}

One particular and surprising benefit of \toolname
  was its handling of inputs that cannot be evaluated
  given the maximum precision cap.
In the prior uniform precision-doubling approach,
  these inputs would require re-evaluating each primitive
  at every precision, up to and including the maximum precision,
  in order to confirm that evaluation was, in fact, impossible.
While such inputs were rare,
  the time needed for this fruitless re-evaluation was so large
  that it was a substantial annoyance to users.
We added an ``early exit'' feature to \toolname
  to address this case.
With early exit, if the precision assigned to a primitive
  is higher than the maximum precision cap,
  \toolname will immediately halt expression execution.
This ``early exit'' can be triggered
  as early as the first tuning iteration,
  radically cutting evaluation time.
Herbie enables the early exit feature by default.

\subsection{Case Studies}

We now illustrate slack, correct rounding, and early exit
  with individual case studies.

\paragraph{Slack mechanism}

% \begin{figure}
% \[
% \begin{array}{llllll}
% z_i & \text{First tuning iteration} & \text{Second tuning iteration} \\\hline
% z_4 = z_3 - z_2
%   & 53 + 3 + 2 = 58
%   & 53 + 3 + 2 = 58 \\
% z_3 = \cos(x)
%   & 58 + 2 + (2 + \slack[1]) = 574
%   & 58 + 2 + (575 + \slack[2]) = 1659 \\
% z_2 = \cos(z_1)
%   & 58 + 2 + (2 + \slack[1]) = 574
%   & 58 + 2 + (575 + \slack[2]) = 1659 \\
% z_1 = x + e
%   & 574 + 2 + (998 + \slack[1]) = 2086
%   & - \\
% \end{array}
% \]

% \caption{
%   Precision assignment in the first and second iteration
%     for the slack mechanism case study.
%   Note that, even though the slack mechanism
%     partially re-introduces search,
%     amplification factors still provide useful information
%     and \toolname still converges to a high precision
%     much more quickly than Baseline.
% }
% \label{fig:slack-case-study}
% \end{figure}

Consider the benchmark $\cos(x) - \cos(x+e)$
  with inputs $x = 10^{100}$ and $e = 10^{-100}$.%
\footnote{This input (and others in the case studies)
  are a simplified form of one of the \NumTunedPoints inputs from RQs 1--3.}
The terms being subtracted differ by a tiny amount,
  and need to be evaluated
  to roughly 120 digits (400 bits) to achieve a correctly-rounded result,
  while the shared addition $x + e$ needs to be evaluated to 240 digits (800 bits)
  of precision.
\toolname reaches a working assignment in four tuning iterations.

After the first, untuned evaluation, \toolname is able to estimate
  the sizes of $x$, $x + e$, and $\cos(x)$.
However, because this first iteration uses low precision,
  $x + e$ covers a wide interval and
  the intervals for $\cos(x+e)$ and the subtraction
  cross zero.
\toolname thus applies the slack mechanism twice,
  to the inputs of the subtraction
  and the $\cos(x+e)$ term.
This results in the following precision assignment:
\begin{enumerate}
\item subtraction does not need to be tuned as it is a root operation and simply requires
  its children to be accurate enough:
  the $P[sub] = 58$ assignment is preserved;
\item each $\cos$ gets tuned by this formula: $P[sub] + 2 + \lceil\ampl_j(sub)\rceil$,
  which for this particular tuning pass is $58 + 2 + (4 + \slack[1]) = 117$ bits,
  where 53~bits is a guess, while the other 64~bits are derived analytically;
\item $x + e$ gets precision of 
  $P[cos] + 2 + \lceil\ampl_1(cos)\rceil = 119 + (335 + \slack[1]) = 507$ bits, 
  with the 106~bits being a guess.
\end{enumerate}

After re-evaluating with these new precision assignments,
  $\cos(x+e)$ no longer crosses zero,
  but the subtraction still does.
Therefore, the second tuning pass applies the slack mechanism once, 
  for the inputs to the subtraction operation.
\toolname then raises:
\begin{enumerate}
\item $\cos(x)$ and $\cos(x+e)$ to $284$ bits, as $P[sub] + 2 + \lceil\ampl_j(sub)\rceil = 60 + (118 + \slack[2])$, with $\slack[2]$ being the only guess, equal to $106$;
\item and $x + e$ to $620$ bits, as $P[cos] + 2 + \lceil\ampl_j(cos)\rceil = 284 + 2 + 334$, with the same guess of 106~bits inherited from the parent.
\end{enumerate}

The third pass still observes subtraction crossing zero.
Therefore, the slack mechanism needs to be triggered one more time.
\toolname raises the precision to
  $556$ bits for both cosine operations;
  and $892$ bits for $x + e$.
This precision assignment yields a correctly-rounded result.
In all, four evaluation iterations are needed, while
  the uniform-precision strategy would need five iterations,
  at $63$, $126$, $252$, $504$, and $1008$ bits.
%1542-optimal
%2062-reval
%4032-baseline
By contrast, the optimal precision assignment for this example
  uses $387$ bits for each cosine,
  $714$ bits for $x + e$,
  and $54$ bits for the subtraction.
\toolname's final assignment is about $5\times$ closer
  to this optimal assignment than the uniform-precision one.
% The main takeaway from this case study is that \toolname does
%   not heavily rely on slack mechanism, it is an important increment, however, in this example
%   it played only about a third part in precision increment.

\paragraph{Correct rounding}
Consider the benchmark $(x+y) \times (z + 1)$
  at the exact double-precision inputs
  $x=1.3002052657264033\times10^{189}$, 
  $y=3.084776002356433\times10^{188}$, 
  and $z=2^{-1000}$.
For this input, all $\ampl_k$ and \intro factors are 1 or less,
  yet after the initial evaluation,
  the output interval is not correctly rounded.
The reason is that these specific $x$ and $y$ values
  are exactly one exponent apart and both end in a 1 bit,
  meaning that their sum, $x + y$,
  lies exactly on a rounding boundary.
The expression thus rounds up only if $z + 1 > 1$
  which requires a thousand bits.
\toolname detects this issue using its correct rounding mechanism,
  increasing the target precision $t$,
  first from $53$ to $53 + \slack[1] = 106$
  then $159$, $265$, $477$, $901$ and finally $1749$
  bits of precision.
Only after the sixth re-evaluation
  is $z + 1$ strictly greater than $1$,
  allowing \toolname to achieve a correctly-rounded result.

\paragraph{Early stopping}
Consider the benchmark $(x+1)^{1/n} - x^{1/n}$
  for $x = 10^{200}$ and $n = 10^{-200}$.
These inputs require raising a large number
  to a large power, causing overflow:
  $(x+1)^{1/n}$ evaluates to
  $[\approx2\cdot10^{323\thinspace228\thinspace496}, +\infty]$
  after the first, untuned evaluation.
The $x^{1/n}$ term likewise overflows,
  and subtracting the two terms yields
  a very wide output interval of $[-\infty, +\infty]$.
The amplification factor for the subtraction is
\[
  \maxlog([2\times10^{323\thinspace228\thinspace496}, +\infty]) - 
  \minlog([-\infty, +\infty]) =  1\thinspace073\thinspace742\thinspace335 + 2\slack[1].
\]
This implies billions of bits of precision
  for the left-hand term, so \toolname,
  correctly, exits early.

\paragraph{Logspan}
Since the $\logspan$ removal optimization does not fire
  on any of our evaluation benchmarks,
  this case study uses a constructed expression.
Consider $x / ((\sqrt{x + 1} - 1) + x^2)$ 
  with input $x = 10^{-100}$.
The value $\sqrt{x + 1} - 1$ has a cancellation,
  meaning its interval spans from 0 to an epsilon value
  until it is assigned a very high precision,
  roughly 690 bits.
Because of this, the denominator has
  a very large $\logspan$.

In its default configuration,
  ignoring $\logspan$s,
  \toolname evaluates this computation with one tuning pass. 
\toolname assigns 58 bits to the division,
  60 bits to the addition of $x^2$,
  665 bits to the subtraction,
  784 bits to $\sqrt{x+1}$ 
  (because the subtraction triggers $\slack[1]$),
  and 785 bits to $x + 1$.
This precision assignment is accurate enough
  to obtain a correctly-rounded output.

% ,
%   but the denominator no longer crosses zero,
%   so in the second tuning iteration,
%   \toolname assigns 231 bits to $\sqrt{x+1}$
%   and 232 bits to $x+1$, this time without using $\slack$.
% This precision assignment results in
%   a correctly-rounded output,
%   but it requires two tuning iterations.

If, instead, $\logspan$ terms were enabled,
  then the first iteration would include
  additional $\logspan$ terms,
  resulting in a higher precision.
Specifically, the division operation's $\ampl_2$
  would compute $2 \logspan(denom)$,
  which would evaluate to roughly 1200 bits.
This substantial precision increase
  would then be propagated through the rest of the denominator,
  resulting in precisions over 2000 bits for some operations.
In total, this precision assignment
  would be four times larger than the default one.
This computation \emph{would} result
  in a correctly-rounded evaluation, 
  but the assigned precisions would be so high
  that evaluation would take about
  20 microseconds, compared with 5 microseconds
  in the default configuration without  $\logspan$ terms.
This explains why \toolname disables $\logspan$ terms by default
  and only re-enables them when proven necessary.

% Notably, the $\logspan$ terms do not stay high
%   through further iterations.
% In fact, if $\logspan$ terms are enabled
%   only for the second tuning iteration,
%   all $\logspan$ terms would then evaluate to $2$ or less,
%   thus resulting in almost no change to the precision assignment.

%\section{Proofs and Derivations}
%\label{sec:proof}
%\input{proof}

\section{Related Work}
\label{sec:relwork}
H.~Boehm was an early advocate
  of evaluating real-number expressions to high precision
  using ``constructive reals''~\cite{boehm-idea}.
A number of implementations of constructive reals exist,
  including in Java~\cite{boehm-java},
  Python~\cite{constructive-real-python},
  Caml~\cite{constructive-real-verified}, and others.
Early computed real implementations used lazy digit sequences,
  but later work~\cite{boehm-fast,boehm-compare}
  showed that Ziv's strategy~\cite{ziv}
  with interval arithmetic is typically faster.
IRRAM~\cite{irram} and RealLib~\cite{reallib},
  in particular, introduced a number of techniques
  that are widely used today;
  in fact, the most widely-used constructive reals library
  is used in Android's Calculator app~\cite{api-for-real-numbers}.
Other prominent libraries include
  Arb~\cite{arb}, Moore~\cite{moore-ivals}, MPFI~\cite{mpfi},
  Rival~\cite{movability}, and others;
  interval arithmetic in particular
  is standardized by IEEE~1788~\cite{ieee-1788}.
Existing systems, however, largely focus on
  efficiently computing primitive operations at a given precision,
  while \toolname instead focuses on assigning precisions
  to those primitives.
Perhaps the most similar to the current work
  was proposed by \citet{constructive-real-tweaks}:
  mixed-precision evaluation using ``weights''.
However, that work expected users to manually set weights,
  whereas \toolname derives
  a non-uniform precision assignment automatically.

Interval arithmetic and constructive reals are closely related
  to research into automatic worst-case error bounds
  for floating-point programs.
One important early system was Salsa~\cite{salsa,salsa-1,salsa-2},
  which used interval arithmetic to bound
  both ranges of variables and also errors,
  a strategy adopted by other work as well~\cite{precisa}.
Initially phrased as an abstract interpretation,
  later papers~\cite{martel-types}
  rephrased the analysis as a type system,
  where types track both range and precision.
Rosa~\cite{rosa} and Daisy~\cite{daisy}
  use a similar analysis except with affine arithmetic,
  which better handles correlated errors.
FPTaylor~\cite{fptaylor}
  introduced the idea of error Taylor series
  for automatic worst case error bound estimation,
  achieving much lower error.
Satire~\cite{satire} then proposed scaling error Taylor series
  using abstraction, algebraic simplification,
  and automatic differentiation,
  and SeeSaw~\cite{seesaw} extended these techniques
  to handle bounded control flow.
Like \toolname, Satire saw higher-order error
  as a computational bottleneck.
In most worst-case error bound tools,
  the goal is achieving the smallest possible sound error bound,
  unlike in this paper, where the focus is primarily on
  computational efficiency.
ATOMU~\cite{atomu} proposed computing
  first-order error from condition numbers.
While this provides an algorithmic speedup,
  ATOMU did not evaluate it;
  instead, ATOMU focused on condition numbers
  as a method for driving input generation.

Automatic precision tuning is also an active area of research,
  starting with Precimonious~\cite{precimonious},
  which used delta-debugging to assign precisions
  while evaluating accuracy on representative inputs.
Later work~\cite{blame-analysis}
  improved scalability to larger programs.
Meanwhile, FPTuner~\cite{fptuner}
  and OpTuner~\cite{optuner}
  performed sound precision tuning
  using error Taylor series to derive
  error models that could be optimized by
  integer linear programming.
This is slower but leads to lower precisions.
Finally, POP~\cite{pop}
  poses precision tuning as
  a (non-integer) linear programming problem,
  allowing it to be solved much more efficiently,
  though in some settings a ``policy iteration'' technique
  is necessary to achieve an optimal precision tuning.
The linear programming formulation is
  quite similar to \toolname's precision tuning algorithm,
  though POP makes no connection to properties of
  the underlying mathematical operations,
  and as a result POP does not extend to
  various transcendental functions.
Moreover, instead of linear programming,
  which is still quite slow,
  \toolname's precision tuning is fast
  and provably runs in linear time.

\section{Conclusion}
\label{sec:conclusion}
Current algorithms for evaluating real expressions to high precision
  use a uniform precision for all operations.
This wastefully allocates too much precision to most primitives.
We instead introduce a fast, sound method
  for assigning precisions to primitives
  so as to compute correctly-rounded results
  for real expression evaluation.
Our method leverages information
  from low-precision evaluations
  to determine the correct precision to evaluate each primitive at.
\toolname evaluates real expressions
  $\RivalAvgSpeedupOverSollya\times$ faster
  than the state-of-the-art Sollya tool,
  with \toolname's advantage rising to $\RivalMaxSpeedupOverSollya\times$
  for the most difficult input points.
An integration into the widely-used Herbie numerical compiler
  shows that \toolname's advantage carries over
  into real-world use cases,
  with additional advantages for difficult inputs.

%\appendix

\section{Disclosure of Use of AI in Research}
\label{sec:ai_use}
OpenAI ChatGPT and Codex, Anthropic 
  Claude and Claude Code were used 
  in all aspects of \toolname development in the Rust language.
Originally, \toolname was fully written in 
  the Racket language
  without use of AI,
  and then translated to Rust with 
  significant AI help, human supervision and testing 
  against the original Racket version.
Additionally, AI was used in 
  writing some parts of evaluation 
  scripts, such as the Ziv machine, 
  scripts for generating \Cref{fig:density_cdf} and 
  \Cref{fig:optimality_plot}; styling of all the 
  figures presented in \Cref{sec:eval}; 
  help in writing \Cref{algo:evaluation_loop} and 
  \Cref{algo:tuning_algo}.

% OpenAI ChatGPT and Codex were used
%   in all aspects of \toolname development,
%   including the specification language,
%   implementation language,
%   solver engine, and evaluation cases,
%   as discussed in the paper.
% Additionally, ChatGPT was used
%   to generate tables in this work,
%   including \Cref{tab:fp-types}
%   and editing of \Cref{tab:verification-time}.
% ChatGPT-generated tables were audited by the authors
%   to ensure correctness.

\section{Data Availability Statement}
\label{sec:data-avail}
\toolname is a program implemented
  in the Rust language.%
\footnote{Development continues at GitHub: \url{https://github.com/herbie-fp/rival3}}
The program has a command-line interface
  where examples from \Cref{sec:overview} can be  reproduced
  and \toolname's precision assignments can be viewed.
Scripts are available
  for reproducing the results from \Cref{sec:eval},
  including the exact input points used in this paper,
  though the speedup may vary with the machine used.

For the artifact evaluation we
  provide%
\footnote{Zenodo artifact link: \url{https://zenodo.org/records/23203478}}
  the \toolname program,
  instructions for installation of Sollya,
  instructions for running the experiments
  and the scripts for reproducing all plots and results.
The experiments take about 4.5 hours to run,
  and the scripts produces all the plots
  in the paper and all metrics presented in it.
The experiments require Ziv, Ziv+ and Sollya.
Ziv and Ziv+ are variations of \toolname and provided with it,
  but Sollya 8.0 needs to be installed from its website;
  as of October 25 2026, this website is accessible.
The provided scripts handle all interaction with Sollya,
  written by us.
The integration with Herbie, however,
  is not included in the artifact,
  as the Herbie numerical compiler is a moving target
  and we anticipate needing to make additional changes,
  both to it and to \toolname.

\bibliographystyle{ACM-Reference-Format}
\bibliography{references}

\end{document}